\documentclass[preprint2]{aastex63}
\usepackage{graphicx,tabularx,natbib}
\hypersetup{linkcolor=red,citecolor=blue,filecolor=cyan,urlcolor=magenta}

\shorttitle{J1328+2751}
\shortauthors{Nandi et al.}

\begin{document}

\title{Double-peaked Lines, Dual VLBI Components, and Precessing Jets in J1328+2751}
\correspondingauthor{S. Nandi}
\email{snandi@ncra.tifr.res.in}
\author[0000-0002-0786-7307]{S. Nandi} 
\affiliation{National Centre for Radio Astrophysics - Tata Institute of Fundamental Research,
S. P. Pune University Campus, Post Bag 3, Ganeshkhind, Pune 411007, India}\\

\author[0000-0001-9707-3895]{A. Caproni}
\affiliation{N\'ucleo de Astrof\'\i sica, Universidade Cidade de S\~ao Paulo R. Galv\~ao Bueno 868, Liberdade, S\~ao Paulo, SP, 01506-000, Brazil}

\author[0000-0003-3203-1613]{P. Kharb}
\affiliation{National Centre for Radio Astrophysics - Tata Institute of Fundamental Research,
S. P. Pune University Campus, Post Bag 3, Ganeshkhind, Pune 411007, India}

\author[0000-0001-8428-6525]{B. Sebastian}
\affiliation{National Centre for Radio Astrophysics - Tata Institute of Fundamental Research,
S. P. Pune University Campus, Post Bag 3, Ganeshkhind, Pune 411007, India}

\author[0000-0002-9711-6207]{R. Roy}
\affiliation{The Inter-University Centre for Astronomy and Astrophysics, S. P. Pune University Campus, Ganeshkhind, Pune 411007, India}

\begin{abstract}
In this work, we report a radio galaxy with precessing jets, double-peaked emission lines and the presence of two compact radio components with a projected separation of $\sim$6~parsec in Very Long Baseline Interferometry (VLBI) observations. The emission line peak separations could be suggesting the presence of a supermassive black hole binary (BBH) with a separation of 6.3 parsec, matching the VLBI results. The kinematic jet precession model applied to the jets of J1328+2751 indicates that if it is a BBH, the accretion disk of the primary black hole is not coplanar with the binary system orbit, making its jet precess under the effect of the torque produced by the secondary black hole.  However, we find that the Bardeen-Petterson effect can also provide precession timescales compatible with the jet precession period inferred in this source. This source has previously been identified as a restarted double-double radio galaxy (DDRG). Our findings therefore have important ramifications for the nature of DDRGs in general.
\end{abstract}

\keywords{black hole physics --- galaxies: active --- galaxies: individual (J1328+2752) --- galaxies: jets --- radio continuum: galaxies --- techniques: interferometric}

\section{Introduction}\label{sec:intro}
While the existence of supermassive black hole binaries (BBHs) is predicted by hierarchical galaxy formation, the number of candidate BBHs is only a handful \citep[see the compilation in][and references therein]{Rubinur2018}.
The confirmation of BBH candidates typically requires multiple signatures through multi-wavelength data using X-ray, optical and radio telescopes. However, the resolving power of these telescopes, which is the ability to clearly delineate features in astrophysical objects, is vastly varied. Very Long Baseline Interferometry (VLBI) is the only technique that can identify radio emission from accreting BBHs at projected separations of a milli-arcsecond, which correspond to parsec-scales for nearby active galactic nuclei (AGN). VLBI has been able to identify a $\sim$7.3 parsec BBH in a nearby radio galaxy \citep{Rodriguez06}, a $\sim$0.35 parsec BBH in a nearby Seyfert galaxy \citep{Kharb17}, and a $\sim$138 parsec BBH in a distant quasar \citep{Deane14}. While some AGN show large changes in their jet directions distorting them into S- or Z-shapes \citep[regarded as an indirect signature of the presence of BBHs;][]{1980Natur.287..307B, 1982ApJ...262..478G, Deane14, Kharb17} and others show the presence of double-peaked emission lines in their optical spectra \citep[regarded as another indirect signature of BBHs;][]{2009ApJ...698..956C,Rubinur2017}, seldom does a single source show all these characteristics.

\begin{figure*}
\includegraphics[width=8cm,trim=70 0 0 0]{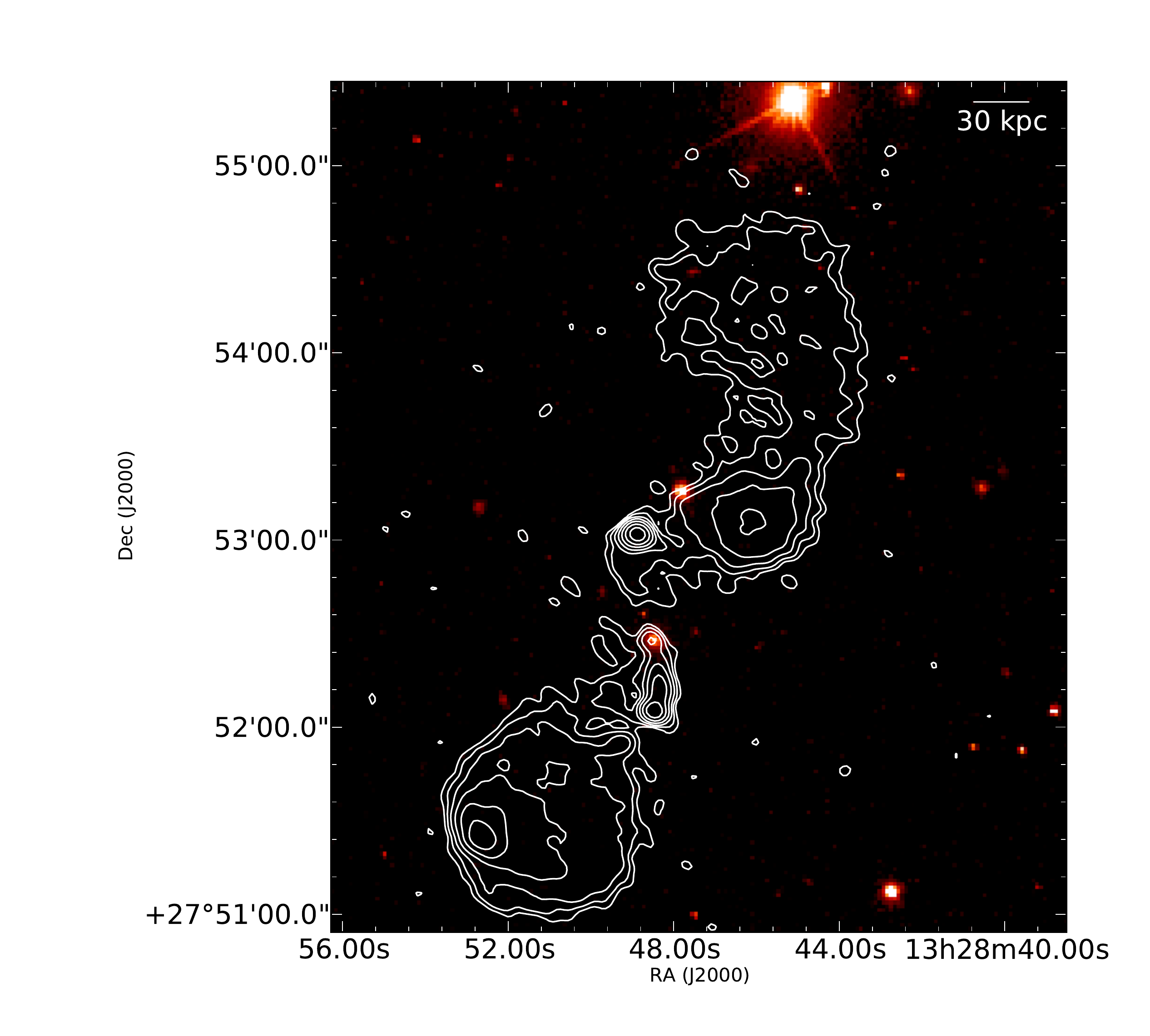}
\includegraphics[width=8cm,trim=0 0 250 0]{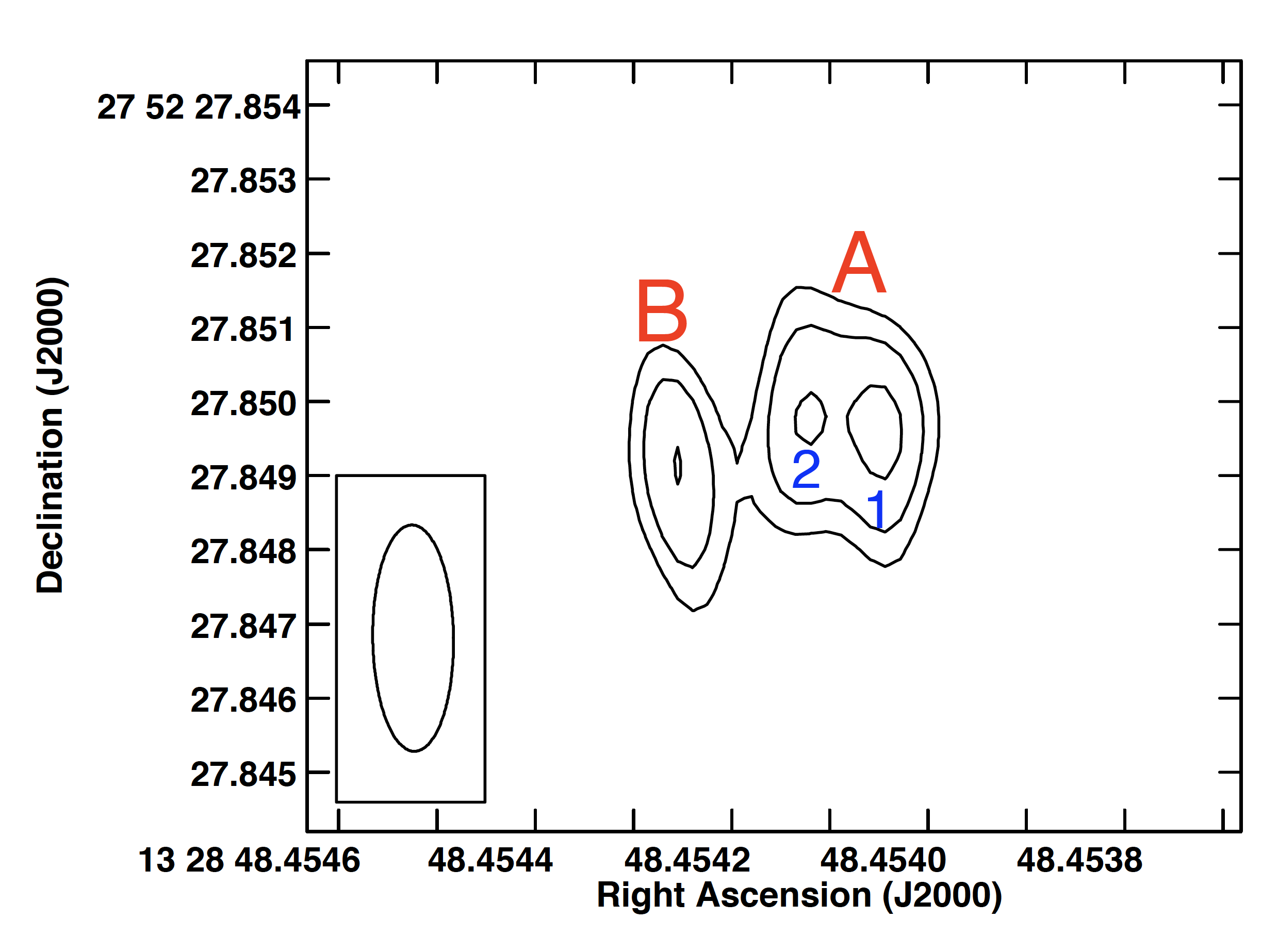}
\caption{Left panel: GMRT 607 MHz radio contour of IRAS J1328$+$2752 is overlaid on the SDSS r-band image. The contour levels start at 3$\sigma$[-1, 1, 2, 4, 8, 16, 32, 64,], where rms noise $\sigma$=0.07 mJy/beam.
Right panel:4.98 GHz contour image of J1328$+$2752 from the VLBA. The contour levels are in percentage of peak surface brightness (=170.5 $\mu$Jy~beam$^{-1}$) and increase in steps of $\sqrt{2}$ with the lowest level being $\pm45$\%. The projected separation between components A (core-jet structure with sub-components 1 and 2) and B is $\sim$6 parsec.}
\label{figureGMRT}
\end{figure*}

J1328+2752 has been classified as `double-double radio galaxy' (DDRG) in the literature \citep{2000MNRAS.315..371S}. At a redshift of $z=0.0911$ and a luminosity distance of {412 Mpc (assuming a $\Lambda$CDM cosmology with $H_0$=71 km s$^{-1}$ Mpc$^{-1}$, $\Omega_{\rm m}$=0.27 and $\Omega_{\rm vac}$=0.73)}, it is one of nearby DDRG, affording us a close look at its nucleus. It contains two pairs of radio jets and lobes launched from its accreting supermassive black hole (SMBH) - accretion disk system, referred to as an AGN. Unlike typical DDRGs however, the two pairs of radio jets/lobes are highly misaligned with each other {($\sim30^\circ$; \citealt{Nandi17})}. The outer radio lobes ($\sim$413~kpc in total projected extent) in J1328+2752, and the inner radio lobes ($\sim$96 kpc in total projected extent) are asymmetric in morphology (Figure~\ref{figureGMRT}, left panel). The radio ``core'' comes into prominence at progressively higher radio frequencies, indicating its inverted radio spectrum. It is not clearly detected at 325 MHz but becomes visible at 610 MHz in the Giant Metrewave Radio Telescope (GMRT) images as well as the 1.4~GHz Very Large Array (VLA) Faint Images of the Radio Sky at Twenty centimeters (FIRST) image \citep{2019MNRAS.486.5158N}. 

J1328+2752 is hosted by a giant elliptical galaxy which are thought to be built from multiple prior galaxy mergers \citep[][see Appendix \ref{sec:SDSS} for further details]{1972ApJ...178..623T, 2017ARA&A..55...59N, 2020MNRAS.498.2766W}. Each merger event should have deposited one or more SMBHs in their centres considering the fact that all galaxies host SMBHs in their centres \citep{Harms94,Genzel97}. 

The Sloan Digital Sky Survey (SDSS) optical spectrum of J1328+2752 shows double peaks in the [Ne III] $\lambda3869$, H$\gamma$ $\lambda4342$, H$\beta$ $\lambda4863$, [O III] $\lambda4959$, [O III] $\lambda5007$, [O I] $\lambda6300$, H$\alpha$ $\lambda6563$, [N II] $\lambda\lambda6548,6584$, and [S II] $\lambda\lambda6718,6732$ emission lines. The average velocity separation of the two line components is \citep{Nandi17} $\Delta v = 235.0\pm10.5$ km s$^{-1}$. This velocity separation can be produced by the presence of two SMBHs, each with their corresponding line-emitting gas regions \citep{Komossa06,Kharb2020}. Indeed, the Baldwin, Phillips and Terlevich (BPT) diagnostic diagram shows AGN characteristics for both the emission-line components \citep{Nandi17}. However, double-peaked emission lines can also be produced  due to narrow line region (NLR) gas kinematics, outflows and jet medium interaction \citep{2010ApJ...708..419C, 2011ApJ...739...44R, Kharb2019} or NLR gas in rotating disks \citep{2012ApJ...752...63S, 2012ApJ...753...42C,Kharb15, 2019MNRAS.484.4933R}. 

\section{Data and Results}
\subsection{J1328+2752 on parsec scales} \label{sec:VLBA}
The VLBI data were acquired with 9 antennas (minus the Pie Town antenna) of the Very Long Baseline Array (VLBA) at 4.98 GHz (Project ID BS267B) on August 11, 2018 in a phase-referenced experiment. The total bandwidth of the data was 256 MHz. The calibrators DA193 and J1800+3848 were used as the fringe-finders for the experiment. J1333+2725 was used as the phase calibrator. The data were reduced using the VLBARUN data reduction pipeline with appropriate parameters inside AIPS, with Los Alamos (LA) as the reference antenna. No self-calibration was carried out.

The VLBI image shows a core-jet structure with a PA of 80$^\circ$ (component A with sub-components 1 and 2 in Figure~\ref{figureGMRT}, right panel) and another single component (B) separated in projection by $\sim$6 parsec. The final r.m.s. noise in the image was 23~$\mu$Jy~beam$^{-1}$, while the synthesized beam was $3\times1$ mas at a PA of 0.72$^\circ$. Using the Gaussian fitting AIPS task JMFIT, we found that the peak surface brightness of components 1 and 2 are 171~$\mu$Jy~beam$^{-1}$ and 142~$\mu$Jy~beam$^{-1}$, respectively, while for component B is 156~$\mu$Jy~beam$^{-1}$. 

Using the relations from \citet{Ulvestad05}, the brightness temperature $T_B$ of the unresolved components 1 and 2 in A are 1.6$\times10^7$~K and $1.0\times10^7$~K, respectively, while $T_B$ of component B is $1.1\times10^7$~K. The high brightness temperatures indicate that all three components are non-thermal sources related to AGN activity. It suggests that components A and B could be the unresolved bases of jets from two accreting SMBHs.
However, the in-band spectral indices cannot be accurately determined due to the narrow bandwidth of the acquired data. Thus, multi-frequency VLBI data are needed to identify inverted/flat spectrum radio cores in the image. Multi-epoch VLBI observations can also look for proper motion in the VLBI components, which can also confirm the stationary/moving components as core/jet. Such observations have been proposed for, and will be presented in a forthcoming paper, pending their acceptance.

Using Kepler's third law, we can estimate the distance between the two putative SMBHs, $d_\mathrm{BH} = \frac{GM_\mathrm{tot}}{\Delta v^2}$, where $G$ is the gravitational constant and $M_\mathrm{tot}$ 
is the total mass of the BBH system. In the case of J1328+2752, $\log\left(M_\mathrm{tot}/M_\odot\right) = \left(7.91\pm0.17\right)$, estimated by \citet{Nandi17} from the stellar velocity dispersion of the host galaxy, $\sigma_\ast=176.8\pm9.7$ km s$^{-1}$, and the well known $M_{BH}-\sigma_\ast$ relation \citep{Ferrarese00}. This leads to $d_\mathrm{BH} = 6.3^{+3.1}_{-2.1}$ parsec (uncertainties at 1$\sigma$ level), which corresponds to an orbital period in the observer's reference frame of $P_\mathrm{orb,obs}=0.17^{+0.13}_{-0.09}$~Myr. Moreover, our analyses suggest that this putative BBH in J1328+2752 would be gravitationally bound and Keplerian (see Appendix \ref{sec:SMBBSKepler} for further details).

\subsection{J1328+2752 on Kpc-scales}\label{sec:jet_model}

The case of a BBH is further bolstered by the kpc-scale morphology of the source. The low-frequency radio images of J1328+2752 show an S-shaped brightness distribution, which has been observed in other radio galaxies as well \citep{1985A&AS...59..511P, 1988ApJ...329..122H}.
Several mechanisms can explain the twisted jet morphology. Jets may follow bent S-shaped trajectories due to the strong interactions with magnetized surrounding medium or dense interstellar medium \citep[e.g.,][]{Lehnert99,2018MNRAS.480.3644R}.
Two main magnetohydrodynamic instabilities- Kelvin-Helmholtz (KH) and current driven (CD) kink instability play a significant role in jet geometry and stability \citep{Hardee87,2020MNRAS.499..681M}. 
Jet structure can also be affected by ram pressure or buoyancy force \citep{Blanton01,2019AJ....157..126G}. The relative velocity of the associate galaxy is the main mechanism of formation of ram pressure while the buoyancy force arise due to the density gradient between jet and the surrounding medium. Alternatively, the other possible explanation is that the jet/counterjet system may have been subject to precession
\citep[e.g.,][]{1978Natur.276..588E,Gower82,Kharb2006,Kharb14,Rubinur2017},
which is also corroborated by hydrodynamic simulations of jets
(e.g., \citealt{2010ApJ...713L..74F,2016MNRAS.458..802N}). In this work, we have explored the feasibility of such a phenomenon in J1328+2752 adopting a pure kinematic precession model used previously in the context of blazars (e.g., \citealt{2000A&A...355..915A, 2004ApJ...602..625C, 2013MNRAS.428..280C}).

\subsubsection{Kinematic precession model}\label{sec:Kinematic_model}
Our precession model assumes that jet and counter-jet are formed by plasma elements with a constant speed $\beta c$, where $c$ is the speed of light and $0\leq\beta<1$. Due to the precession motion, these plasma elements are ejected from the core region along different directions on an apparent cone surface with a semi-aperture angle $\varphi_0$ and an axis that forms an angle $\phi_0$ with the line of sight. After an interval corresponding to the precession period measured at the observer's reference frame, $P_\mathrm{prec,obs}$, both jet and counter-jet will point again to the same initial direction. The successive ejections of the plasma elements will form a helical path on the precession cone that, when project onto the plane of the sky, will produce bent trajectories around a fixed direction (the projection of the precession cone axis) that forms an angle $\eta_0$ in relation to the North (positive from North to East). The two remaining free model parameters are the precession phase, $\tau_\mathrm{0,s}$, and the sense of precession, $\iota$ ($\iota=1$ for clockwise sense and $\iota=-1$ for counter-clockwise sense \citep{2009MNRAS.399.1415C}.

Unfortunately, no direct measurements of the jet speed $v=\beta c$ and the jet viewing angle $\phi$ exist in the literature for J1328+2752. Thus, we derived limits for their values using the arm-length-ratio, $R_\mathrm{len}$, and the flux density ratio, $R_\mathrm{flux}$, \citep{1997ApJ...485L...9T}:

\begin{equation} \label{Rlen}
R_\mathrm{len} = \frac{L_\mathrm{j}}{L_\mathrm{cj}}, 
\end{equation}

\begin{equation} \label{Rflux}
R_\mathrm{flux} = \frac{S_\mathrm{\nu,j}}{S_\mathrm{\nu,cj}}, 
\end{equation}
\\where $L_\mathrm{j}$ and $L_\mathrm{cj}$ are, respectively, the lengths of the jet and counter-jet measured from the radio core, while $S_\mathrm{\nu,j}$ and $S_\mathrm{\nu,cj}$ are, respectively, the measured flux densities at a frequency $\nu$ for the jet and counter-jet ($S_\nu \propto \nu^{-\alpha}$, where $\alpha$ is the spectral index). 

As the jet length is proportional to the apparent speed, $\beta_{\mathrm{obs}}$,

\begin{equation} \label{betaobs}
\beta_{\mathrm{obs}} = \frac{\beta\sin\phi}{1-\beta\cos\phi},
\end{equation}
\\ it is possible to estimate $\beta\cos\phi$ from

\begin{equation} \label{betacosphiRlen}
\left(\beta\cos\phi\right)_{R_\mathrm{len}} = \left(\frac{R_\mathrm{len}-1}{R_\mathrm{len}+1}\right). 
\end{equation}

Similarly, as the observed flux density is proportional to $\delta^p = \gamma^{-p}\left(1-\beta\cos\phi\right)^{-p}$, where $p=2+\alpha$ for a continuous jet \citep{1985ApJ...295..358L}, an additional estimate for $\beta\cos\phi$ can be obtained from

\begin{equation} \label{betacosphiflux}
\left(\beta\cos\phi\right)_{R_\mathrm{flux}} = \left(\frac{R_\mathrm{flux}^{1/p}-1}{R_\mathrm{flux}^{1/p}+1}\right). 
\end{equation}

\begin{figure*}
\centering{
\includegraphics[width=15cm]{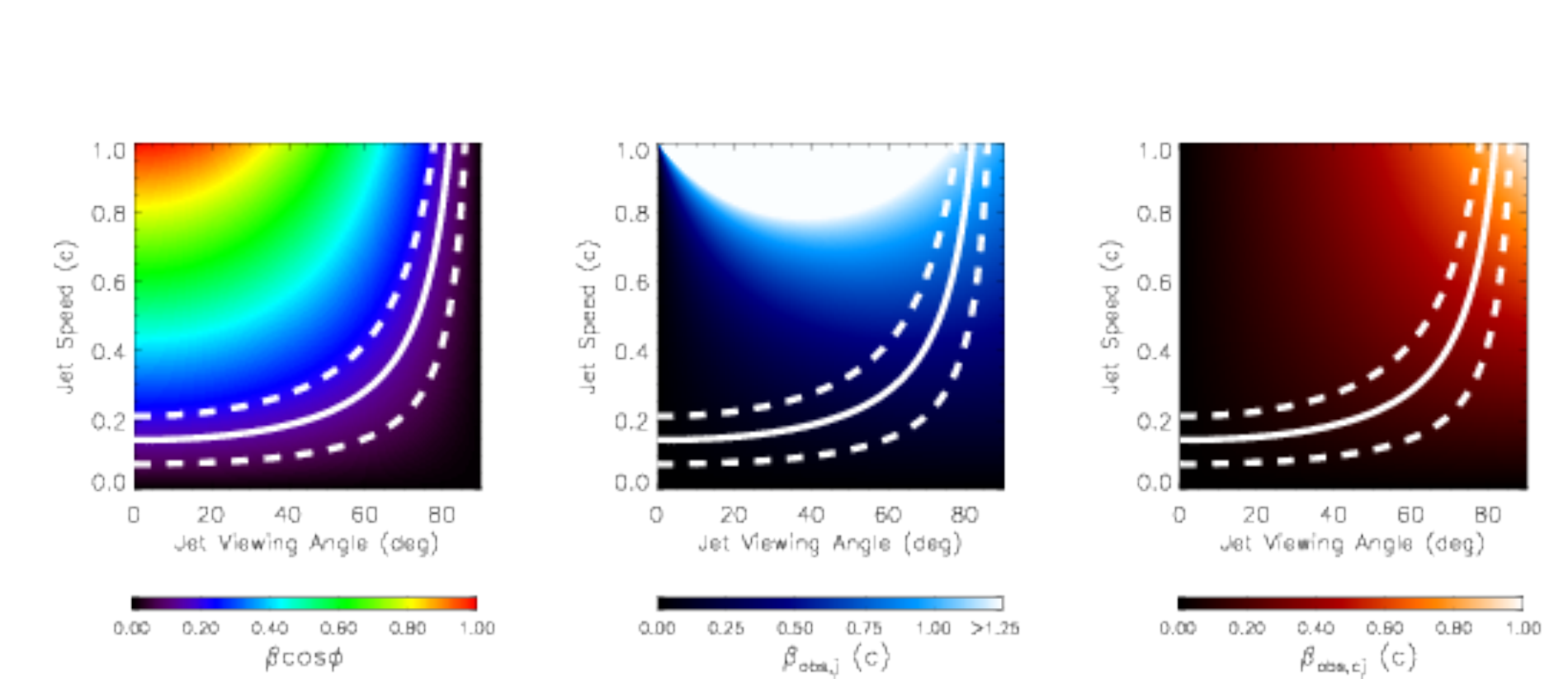}}
\caption{{\it Left:} Two-dimensional behaviour of $\beta\cos\phi$ as a function of $\beta$ and $\phi$ (coloured degrade). Solid white line represent the jet speed as a function of $\phi$ derived from the weighted mean of $\beta\cos\phi$ involving all the estimates for $R_\mathrm{len}$ and $R_\mathrm{flux}$. Dashed white lines refer to 3$\sigma$-level estimates. {\it Middle:} Two-dimension behaviour of $\beta_\mathrm{obs}$ for the jet a function of $\beta$ and $\phi$ (coloured degrades). {\it Right:} Same plot but for the counter-jet. } \label{betacosphi_betaobsjet_cjet_plot}
\end{figure*}

The lengths and flux densities of the jet and counter-jet of J1328+2752 were estimated from the modelling of our three arc-second-scale radio images, using a superposition of elliptical Gaussian components. We adopted the outermost regions of the jet and counter-jet as a proxy for such determinations, with the jet corresponding to the structure seen in the northern part of our radio images. The estimates for $R_\mathrm{len}$ and $R_\mathrm{flux}$, as well the respective values for $\beta\cos\phi$ are shown in Table \ref{Rlen_Rflux_betacosphi}, implying a weighted mean value for $\beta\cos\phi = 0.143\pm 0.023$. This value was used to derive the jet speed as a function of the jet viewing angle, as is shown in the left panel of Figure \ref{betacosphi_betaobsjet_cjet_plot}. A very conservative lower limit for $\beta$ is obtained imposing $\phi=0$, which leads to $\beta > 0.07$ at 3$\sigma$-level, besides an upper limit for $\phi$ taking $\beta\rightarrow 1$, implying $\phi < 86^\circ$. 

\begin{deluxetable*}{ccclD}
\tablenum{1}
\tablecaption{Derived values for arm length and flux density ratios, as well as for $\beta\cos\phi$ at 0.325, 0.61 and 1.42 GHz. The uncertainties in each parameter correspond to the 1$\sigma$ level.\label{Rlen_Rflux_betacosphi}}
\tablewidth{0pt}
\tablehead{
\colhead{$\nu$ (GHz)}& \colhead{$R_\mathrm{len}$} &  \colhead{$R_\mathrm{flux}$} &
\multicolumn2c{$(\beta\cos\phi)_{R_\mathrm{len}}$ } & \colhead{$(\beta\cos\phi)_{R_\mathrm{flux}}$}$\star$
}
\decimalcolnumbers
\startdata
1.420  &  1.356  $\pm$   0.082  &   1.946  $\pm$    0.373   &   0.151  $\pm$    0.029     &     0.129  $\pm$    0.037     \\
0.610  &  1.441  $\pm$   0.049  &   3.041  $\pm$    0.367   &   0.181  $\pm$    0.016     &     0.214  $\pm$    0.022     \\
0.325  &  1.439  $\pm$   0.058$\dagger$ &  1.744  $\pm$    0.111  &     0.180  $\pm$    0.019   &     0.108  $\pm$    0.012 \\ 
\enddata
\tablecomments{$\star$ We have adopted $\alpha = 0.56$ in the calculations, obtained from the average of the spectral indices for the NW and SE jet components after a power-law fitting of their flux densities at 325, 610 and 1420 MHz. $\dagger$ Calculated from the mean value of the core coordinates at 0.61 and 1.42 GHz due to the non-detection of the core region at 325 MHz by our model-fit. \\}
\end{deluxetable*}

We show also in Figure \ref{betacosphi_betaobsjet_cjet_plot} the general behaviour of $\beta_\mathrm{obs}$ for jet and counter-jet in terms of $\beta$ and $\phi$. The same white curves plotted in its left panel are superposed on each $\beta_\mathrm{obs}$--diagrams to show the expected ranges for the apparent jet and counter-jet speeds in J1328+2752 (region delimited by dashed lines). Allowed solutions are usually subluminal, not exceeding $\beta_{\mathrm{obs}}\sim1.2c$. A minimum value of $\sim 0.013c$ for $\beta_{\mathrm{obs}}$ is obtained considering the estimated total length between south and north outermost regions ($\sim$413 kpc) and the upper limit of about 52 Myr to the age of the particles in those structures \citep{2019MNRAS.486.5158N}, in full agreement with the range of apparent speeds derived in this work.

\begin{figure*}
\centering{
\includegraphics[width=15cm]{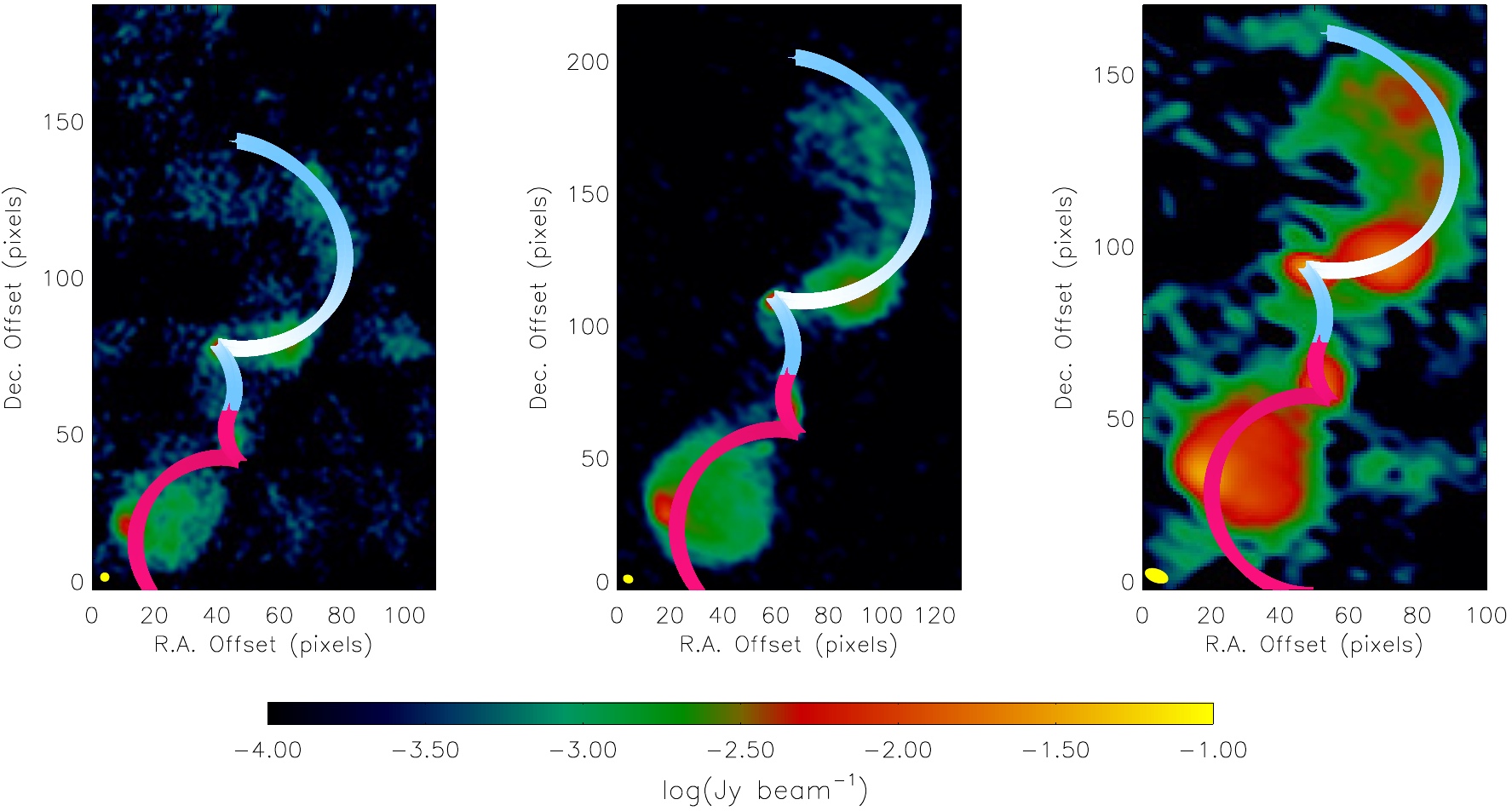}}
\caption{Superposition of the jet and counter-jet precession helices (blue and purple hues, respectively) for a precession period of 4.8 Myr on the interferometric images of J1328$+$2752 at 1.420 GHz (left panel), 0.610 GHz (middle panel), and 0.325 GHz (right panel). Lighter colours used in the precession helices represent higher Doppler boosting factors. The yellow ellipse in the lower-left corner of the individual panels represents the FWHM of the elliptical synthesized CLEAN beam. Note that images have different angular sizes, with one pixel corresponding to $1\farcs67$, $1\farcs20$ and $1\farcs60$ for 1.420, 0.610 and 0.325 GHz, respectively.}
\label{figureVLBA}
\end{figure*}

We show in Figure~\ref{figureVLBA} the images at 610, 325 and 1400 MHz superposed with the precession helices generated by a particular kinematic precession model\footnote{The determination of the free precession model parameters was done without any usage of formal statistical regression procedures, but only imposing $\beta\cos\phi = 0.143\pm 0.069$ in the attempts of recovering the overall jet/counter-jet radio morphology.}. Jet and counter-jet are represented, respectively, by blue and purple curves in this figure, with lighter colours associated with higher Doppler boosting factors (roughly between 1.12 and 1.25 for the jet, and 0.80 and 0.86 for the counter-jet). This precession model respects all observational constraints discussed previously and assumes $P_\mathrm{prec,obs}=4.8$ Myr, $\beta=0.17$, $\phi_0=40^\circ$, $\varphi_0=17^\circ$, $\eta_0=-17^\circ$, $\tau_\mathrm{0,s}=0.88$, and $\iota=1$ (clockwise sense). The first plasma elements show in this figure were ejected about 5.7 Myr ago from the core region. The overall jet/counter-jet morphology is fairly described by this pure ballistic model, arguing in favour of a jet precession scenario for J1328+2752. The inner jet components seen in Figure \ref{figureVLBA} coincides with higher values of the Doppler factor predicted by our precession model. Even though our precession model predicts correctly the projected position of the hotspot seen in the SE lobe, it overestimates the length of the counter-jet, which is not surprising given the own hotspot suggests the presence of a shocked zone that could decelerate and/or deflect the counter-jet. 

Note that the precession model shown in Figure~\ref{figureVLBA} is not unique in the sense of providing similar description of the global radio morphology of J1328+2752. For instance, models with a precession period between $\sim$3 and 8 Myr and jet speed from $\sim$ 0.12$c$ to 0.24$c$ will produce similar precession helices using approximately the same additional free model parameters adopted in Figure~\ref{figureVLBA}. 
Indeed, fair precession models that describe simultaneously the S-shaped jets of J1328+2752 and respect the observational constraints presented previously and in Appendix \ref{sec:JetPrecJ1328} are found using $1.5\la P_\mathrm{prec,obs} (\mathrm{Myr})\la52$, $0.075\la\beta\la0.54$, $7\la \phi_0 (^\circ)\la68$, $3\la\varphi_0 (^\circ)\la30$, $-25\la\eta_0 (^\circ)\la-8$ and $0.78\la\tau_\mathrm{0,s}\la0.99$. Moreover, our analyses indicate the existence of a trend where a precession model with shorter (longer) precession periods need simultaneously higher (lower) jet speeds and larger (smaller) values for $\phi_0$ and $\varphi_0$. It also reinforces the importance of an independent and more accurate estimate of the jet speed to narrow the ranges of the precession model parameters provided in this work.

\section{Some physical mechanisms for driving jet precession}\label{sec:PhyMech}

\subsection{Misaligned accretion disk with the orbital plane of the BBH}\label{sec:Non-CoplanarOrb}

The question that arises is whether jet precession inferred from the arcsecond-scale images of J1328+2752 is compatible with the putative parsec-scale BBH. In this section, we analysed the possibility that jet precession is induced by torques in the primary accretion disk due to the secondary black hole in a non-coplanar orbit around the primary one
\citep{1988ApJ...325..628S, 1997ApJ...478..527K}.
Assuming that jet precession is driven by a secondary black hole with mass $M_\mathrm{s}$ in a circular orbit of radius $d_\mathrm{BH}$ around a primary black hole with mass $M_\mathrm{p}$, we can establish a relationship between $P_\mathrm{prec,obs}$ and $P_\mathrm{orb,obs}$ through \citep{2000MNRAS.317..773B, 2017ApJ...851L..39C}


\begin{equation}\label{PprecSMBBHS}
\left[K(s)\cos\varphi_0\right]\left(\frac{P_\mathrm{prec,obs}}{P_\mathrm{orb,obs}}\right)=\left[\frac{\left(1+q_\mathrm{BH}\right)^{1/2}}{q_\mathrm{BH}}\right]\left(\frac{R_\mathrm{out,p}}{d_\mathrm{BH}}\right)^{-3/2},
\end{equation}
\\where $K(s)\approx 0.19-0.47$ adopting a power-law surface density accretion disk with an index $s=-2$ and 0, respectively \citep{1996MNRAS.282..597L, 2000MNRAS.317..773B}, $q_\mathrm{BH}=M_\mathrm{s}/M_\mathrm{p}$, and $R_\mathrm{out,p}$ is the outer radius of the primary accretion disk.

Assuming that $R_\mathrm{out,p}$ corresponds to the tidal radius of the BBH (e.g., \citealt{1977MNRAS.181..441P, 2000A&A...360...57R, 2017ApJ...851L..39C}), we can write $R_\mathrm{out,p}/d_\mathrm{BH}\cong 0.88 f(q_\mathrm{BH})$, where $f(q_\mathrm{BH})$ is defined as \citep{1983ApJ...268..368E}

\begin{equation}\label{fqBH}
f(q_\mathrm{BH})=\left(0.49q_\mathrm{BH}^{2/3}\right)\left[0.6q_\mathrm{BH}^{2/3}+\ln{\left(1+q_\mathrm{BH}^{1/3}\right)}\right]^{-1}. 
\end{equation}

\begin{figure}
\centering{
\includegraphics[width=8cm]{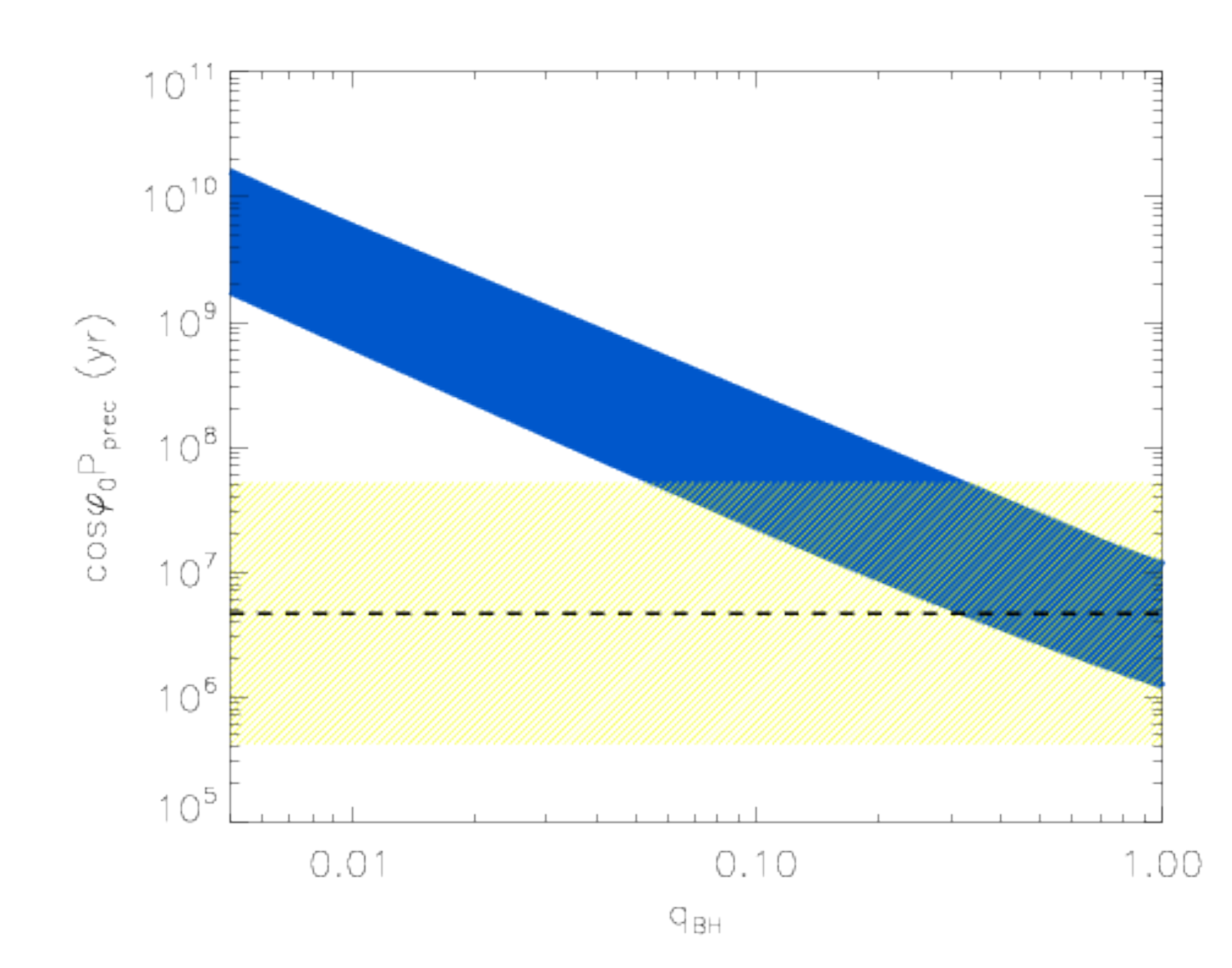}}
\caption{Behavior of $\cos\varphi_0 P_\mathrm{prec,obs}$ as a function of $q_\mathrm{BH}$ in J1328+2752 due to precession of the primary's accretion disk induced by a secondary black hole in a non-coplanar orbit around the primary one (blue stripe) considering 1$\sigma$-level estimate. Hatched yellow rectangle delineates the lower and upper conservative limits for $\cos\varphi_0 P_\mathrm{prec,obs}$ expected for J1328+2752 (see Appendix \ref{sec:JetPrecJ1328}). Dashed horizontal line refers to the jet precession model shown in Figure \ref{figureVLBA}.}
\label{figurecosvarphi0Pprec_vs_qBH}
\end{figure}

We show in Figure \ref{figurecosvarphi0Pprec_vs_qBH} the behaviour of $\cos\varphi_0 P_\mathrm{prec,obs}$ as a function of $q_\mathrm{BH}$ considering the uncertainties in $P_\mathrm{orb,obs}$ and the probable range of $K(s)$. It is also displayed in the same plot a conservative range for $\cos\varphi_0 P_\mathrm{prec,obs}$ ($\sim 0.4 - 52$ Myr) derived from the observational constraints discussed in Appendix \ref{sec:JetPrecJ1328}. The compatibility between the BBH scenario and jet precession is found in the intersection region between blue and yellow stripes, which implies $M_\mathrm{s}\gtrsim 0.05M_\mathrm{p}$. The value of $\cos\varphi_0 P_\mathrm{prec,obs}$ for the jet precession model shown in Figure \ref{figureVLBA} is also plotted in Figure \ref{figurecosvarphi0Pprec_vs_qBH}, which is quite compatible with this proposed scenario if $q_\mathrm{BH}\gtrsim 0.3$.

It is usually thought that geometrically thin AGN disks become gravitationally unstable at distances of about 0.01-1 parsec from the accreting SMBH \citep[e.g.,][]{2003MNRAS.339..937G, 2003MNRAS.341..501S, 2007MNRAS.377L..25K}, which would imply that $R_\mathrm{out,p}\la 1$ parsec. Indeed, water maser emission in some AGNs reveals the existence of rotating gas at distances from the accreting SMBH that range from $\sim$0.03 to 1.33 parsec (e.g., \citealt{1995Natur.373..127M,1996ApJ...472L..21G, 2005ApJ...629..719H, 2011ApJ...727...20K}), bringing observational evidence of AGN accretion disks can extend to the parsec-scale distances. Repeating the previous analysis with $R_\mathrm{out,p} = 0.1$ parsec, for instance, $\cos\varphi_0 P_\mathrm{prec,obs}$ predicted by equation (\ref{PprecSMBBHS}) is always higher than about 80 Myr, compatible with the conservative limits shown by the yellow rectangle in Figure \ref{figurecosvarphi0Pprec_vs_qBH} only if a confidence level higher than $\sim1.5\sigma$ is considered. To reconcile such estimates at $1\sigma$ level would be necessary that $R_\mathrm{out,p} \ga 0.8$ parsec, still within the one-parsec size limit usually expected for geometrically thin accretion disks.\\

\subsection{Geodetic precession}\label{sec:GeoPrec}
Another scenario involving BBHs as a driver of jet precession is the geodetic precession \citep{1975ApJ...199L..25B}. Considering that the jet follows the spin direction of one of the SMBHs, jet precession period can be estimated through \citep{2019MNRAS.482..240K} 

\begin{equation} 
\left(\frac{P_\mathrm{prec}^\mathrm{geo}}{\mathrm{Myr}}\right)\cong 124\frac{(1+q_\mathrm{BH})^2}{q_\mathrm{BH}(3q_\mathrm{BH}+4)}\left(\frac{d_\mathrm{BH}}{\mathrm{1 pc}}\right)^{5/2}\left(\frac{M_\mathrm{tot}}{\mathrm{10^9 \mathrm{M}_\odot}}\right)^{-3/2}  
\end{equation}
\\for a circular orbit. The minimum value of $P_\mathrm{prec}^\mathrm{geo}$ is found setting $q_\mathrm{BH}=1$ in previous equation, which turns out to be about $\sim 305$ Gyr in the case of J1328+2752, incompatible with the precession period derived from the morphology of the kiloparsec-scale radio jet. \\

\begin{figure*}
\centering{
\includegraphics[width=17cm]{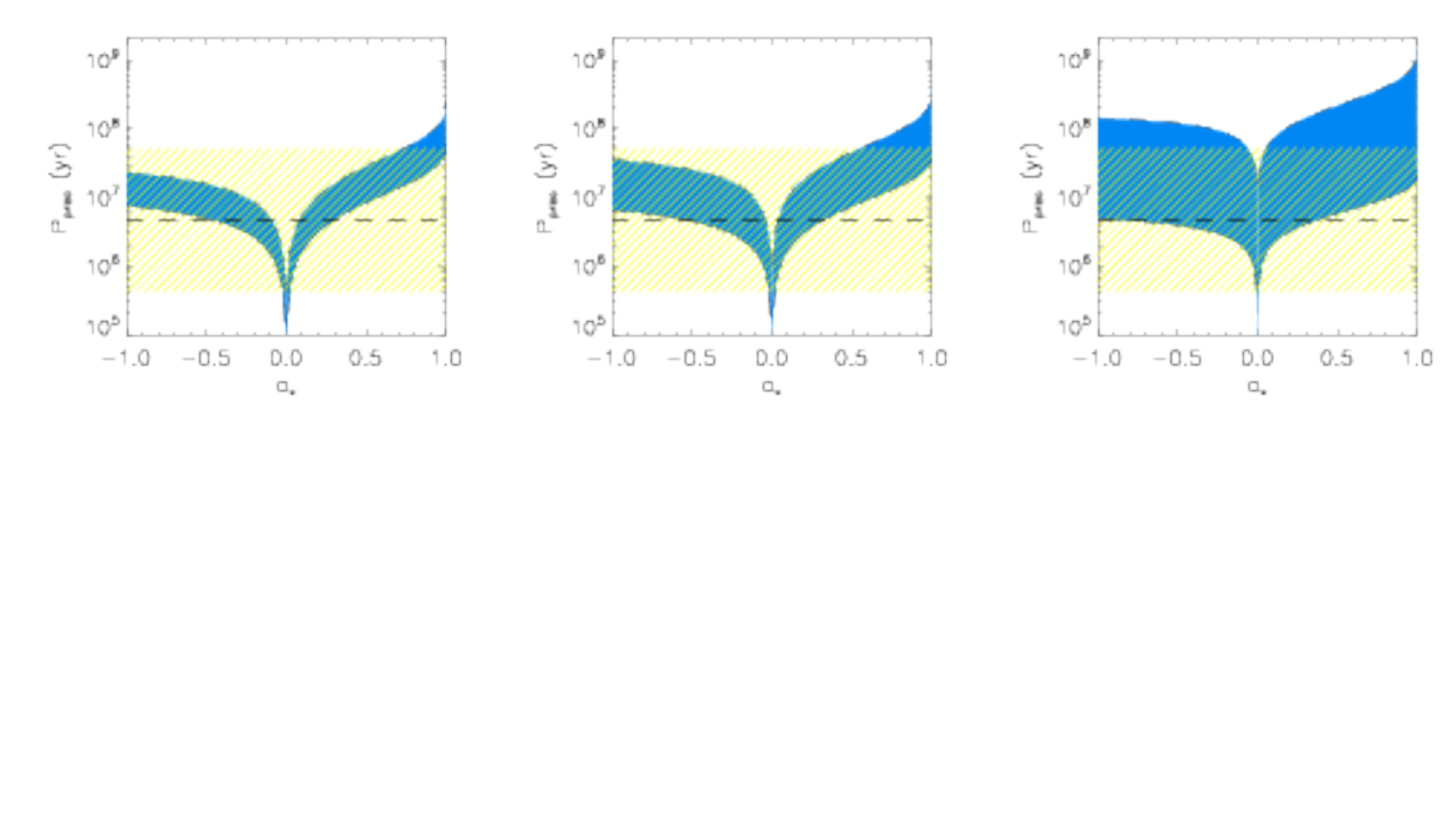}}
\caption{Behavior of the precession period induced by the BP effect as a function of $a_\ast$ in the case of J1328+2752 for $s=-2$ (left panel), -1 (middle panel), and 0 (right panel), and considering $\alpha = 0.1$, $R_\mathrm{out,p}=0.05$ parsec. Upper and lower envelopes of the blue stripes for $P_\mathrm{prec,BP}$ refer respectively to the lower and upper values of $\Sigma_0$. Hatched yellow rectangle delineates the lower and upper conservative limits for $P_\mathrm{prec,obs}$ expected for J1328+2752 (see Appendix \ref{sec:JetPrecJ1328}). Dashed horizontal line refers to the jet precession model shown in Figure \ref{figureVLBA}.}
\label{figureBP}
\end{figure*}

\subsection{The Bardeen-Petterson effect}\label{sec:BP}
In contrast to the previous mechanisms, the Bardeen-Petterson effect (BP; \citealt{1975ApJ...195L..65B}) can induce precession in an accretion disk with no need of a secondary massive body if there is a misalignment between the angular momentum of a spinning black hole, $J_\mathrm{BH}$ and the angular momentum of the accretion disk, $J_\mathrm{d}$. The alignment/counter-alignment between $J_\mathrm{BH}$ and $J_\mathrm{d}$ is forced by a combination between the relativistic Lense–Thirring effect \citep{1918PhyZ...19..156L, 1972PhRvD...5..814W} and the internal viscosity of the accretion disk. The BP effect has been extensively explored in the literature as a potential driver of warping and precession of accretion disks from the point of view of analytical/semi-analytical models (e.g., 
\citealt{1996MNRAS.282..291S, 1998ApJ...506L..97N, 2006ApJ...638..120C, 2006ApJ...653..112C, 2006MNRAS.368.1196L}), as well as through detailed
numerical simulations (e.g., \citealt{2000MNRAS.315..570N, 2007MNRAS.381.1287L, 2007ApJ...668..417F, 2015MNRAS.448.1526N, 2018MNRAS.474L..81L}).

Assuming an accretion disk with surface density, $\Sigma$, shear viscosity $\nu_1$, and vertical shear viscosity $\nu_2$ described by power-law functions of the distance $r$ from the accreting black hole (e.g., \citealt{2006ApJ...638..120C, 2006ApJ...653..112C, 2007MNRAS.381.1617M}), it is possible to estimate the precession period of the jet, $P_\mathrm{prec,BP}$, due to the BP effect as follows

\begin{equation} \label{Pprec_BP}
P_\mathrm{prec,BP} = \cot\left[\frac{\pi\varpi}{2}\right]T_\mathrm{align,BP},
\end{equation}
\\
where $\varpi=1/[2(s+1)]$, $s$ is the power-law index for $\Sigma$, $\nu_1$ and $\nu_2$, 

\begin{equation} \label{SurfDensity}
\Sigma(r) = \Sigma_0 \left(\frac{r}{R_\mathrm{g}}\right)^{-s},
\end{equation}

\begin{equation} \label{nu1}
\nu_1(r) = \nu_{10} \left(\frac{r}{R_\mathrm{g}}\right)^{s},
\end{equation}

\begin{equation} \label{nu2}
\nu_2(r) = \nu_{20} \left(\frac{r}{R_\mathrm{g}}\right)^{s},
\end{equation}
\\where $\Sigma_0$, $\nu_{10}$ and $\nu_{20}$ are constants, and $R_\mathrm{g}=GM_\mathrm{tot}/[c^2(1+q_\mathrm{BH})]$ is the gravitational radius of the accreting black hole. 

For a steady-state accretion disk, the quantity $\nu_{10}$ can be expressed as (e.g., \citealt{1981ARA&A..19..137P})

\begin{equation} \label{nu10}
\nu_{10} \approx \frac{\dot M}{3\pi\Sigma_0},
\end{equation}
\\ where $\dot M$ is the accretion rate of the disk.

The alignment timescale of the accretion disk, $T_\mathrm{align,BP}$, is defined as \citep{2007MNRAS.381.1617M}

\begin{equation} \label{Talign_BP}
T_\mathrm{align,BP} = \frac{c^2}{2\pi\Sigma_0}\left[\frac{G^3M_\mathrm{tot}}{(1+q)}\right]^{-1/2}\frac{T_s}{\cos\left[\frac{\pi\varpi}{2}\right]},
\end{equation}
\\where

\begin{equation} \label{T_BP}
T_s = \left(4\varpi\right)^{2s\varpi} R_\mathrm{g}^{-s\varpi}\left(|a_\ast|c\frac{R_\mathrm{g}^2}{\nu_{20}}\right)^{(1+2s)\varpi}\frac{\Gamma(\varpi)}{\Gamma(1+2s\varpi)},
\end{equation}
\\$a_\ast$ ($|a_\ast|\le1$) is a dimensionless parameter corresponding to the ratio between the actual angular momentum of the black hole and its maximum possible value, $\nu_{20}/\nu_{10} \approx 2(1+7\alpha^2)/[\alpha^2(4+\alpha^2)]$ in the limit of small-amplitude warps \citep{1999MNRAS.304..557O}, and $\alpha$ is the standard viscosity parameter introduced by \citet{1973A&A....24..337S}. 

Following \citet{2006ApJ...653..112C}, we can put lower and upper limits to the value of $\Sigma_0$ based on arguments involving subsonic radial accretion and self-gravitating stability of the disk, respectively. We show in Figure \ref{figureBP} the behavior of $P_\mathrm{prec,BP}$ as a function of $a_\ast$ in the case of J1328+2752 for three representative values of $s$ and assuming $\alpha = 0.1$, $R_\mathrm{out,p}=0.05$ parsec, $q_\mathrm{BH}=0$, and $\ell = L_{\mathrm{bol}}/L_{\mathrm{Edd}} \sim 0.02$\footnote{Following, \citet{2012MNRAS.426..851K}, we have considered $L_{\mathrm{bol}}=C_\lambda\lambda L_\lambda$, where $C_\lambda=9.0$ \citep{2000ApJ...533..631K}, while $\lambda L_\lambda$ is the optical monochromatic continuum luminosity at $\lambda$ = 5100 \AA.}, where $L_{\mathrm{bol}}$ and $L_{\mathrm{Edd}}$ are respectively the bolometric and Eddington luminosities. The BP effect provides precession timescales compatible with the inferred range for the precession period of the kiloparsec-scale jet of J1328+2752 for a wide range of values of $s$ and $a_\ast$. Therefore, jet precession induced by a spinning BH due to the BP effect cannot be ruled out in the case of J1328+2752. 

It is important to emphasize that none of the three precession scenarios analyzed in this work are excluding mechanisms, so that they could be acting simultaneously in J1328+2752. For instance, the coupled evolution of the BH spin and its accretion disk under influence of perturbations of a binary companion orbiting at fixed orbital separation were studied recently by \citet{2020MNRAS.496.3060G}, showing the complex and non-linear interplay between both mechanisms.

\section{Discussion}\label{sec:Discussion}
We note that double-peaked lines can also be produced by line-emitting gas in a rotating disky configuration around a single accreting SMBH \citep{2012ApJ...752...63S, Kharb15, 2019MNRAS.484.4933R}, or bipolar jets from a single AGN pushing the line-emitting gas towards us and away from us, giving rise to the blueshifted and redshifted line components \citep{Kharb2019}.
However, the double peaks are asymmetric in J1328+2752, while they are expected to be symmetric in the case of a disky emission-line gas regions. Moreover, the derived jet speeds are small, making the jet scenario unlikely as well. The double-peaked lines are therefore most likely to be produced in a BBH system. This is an excellent agreement with the VLBI image and the modelling of the kpc-scale jet by precession caused by a BBH. Our results raise interesting possibilities for other DDGRs with highly misaligned lobes (see Appendix \ref{sec:DDRG}). Instead of having multiple activity episodes, they could be sources with parsec- or sub-parsec scale binary separations that cause the jets to precess in an extreme fashion, thereby producing extreme jet/lobe misalignments.

Lastly, there is also a possibility that this galaxy is not a DDRG but a dual jet AGN candidate \citep{palenzula2010}. The dual jet AGN model was initially proposed in an attempt to explain the morphology of the X/S-shaped sources \citep{lal2007}. However, it was argued that the two lobes in X/S shaped sources always appear to show a continuous relationship and do not seem like a superposition of two pairs of lobes selected from a random sample of radio galaxies \citep{saripalli2009}. In almost all the X-shaped sources, the secondary pair of lobes lacks hotspot-like features. Moreover from probability estimates, dual jets are expected to be extremely rare \citep{sebastian2019}. Unlike X/S-shaped sources, J1328+2752 is an extremely rare source in terms of the large-scale morphology which shows two distinct pairs of lobes misaligned with each other. Hence, there exists the possibility that the misaligned lobes are powered by two different jets launched simultaneously by the two different supermassive black holes. Our precession model indicates that if J1328+2752 is a BBH, jet precession could be induced by torques in the primary accretion disk due to the secondary black hole in a non-coplanar orbit around the primary one. 

We note that sources 0402+379 and OJ\,287 show characteristics similar to J1328$+$2752. The period of rotation of 0402+379 is $\sim$1.5$\times10^5$ yr \citep{Rodriguez06} which is comparable to the estimated  orbital period of J1328$+$2751. Maybe the most promising binary pair of SMBHs in blazars, OJ\,287 has a jet precession timescale of $\sim$22 yrs and an orbital period of $\sim$12 yr both measured in the observer's reference frame \citep{2018MNRAS.478.3199B}, suggesting the presence of a much more compact BBH in its nucleus in comparison to the radio galaxies 0402+379 and J1328+2752.

\section{Conclusions} 
We have carried out VLBI 5 GHz imaging and kinematic precession modeling of a radio galaxy J1328+275. The source shows double peaked emission lines spectra in SDSS spectrum. VLBA reveals two radio cores with brightness temperatures of $\sim10^7$~K at a separation of $\sim6$~parsec. The estimated binary separation obtained from the double-peaked lines matched exactly with the VLBI data. Both emission lines and parsec scale image strengthen the feasibility of the BBHs scenario in this source. The precession helices generated by the kinematic model match well with the radio jet at 325, 610 and 1400 MHz. We estimated the jet precession period as roughly between 1.5 and 52 Myr, the precession cone semi-aperture angle between $\sim3\degr$ and $30\degr$ with a line of sight inclination angle of between $\sim 7\degr$ and $68\degr$.
Our results also indicate that jet precession in J1328+275 may be induced by the Bardeen-Petterson effect (induced by a Kerr SMBH with its angular momentum vector misaligned with that of its accretion disk), or by torques in the primary accretion disk due to the secondary black hole in a non-coplanar orbit around the primary one. However, spectral index information is essential to confirm the BBH scenario of this system. We have already proposed for VLBI observations which 
will be deep enough to probe the structures in detail as well as estimate the apparent motion in the parsec-scale jet. 

\acknowledgements
We would like to thank the anonymous referee for the  
constructive comments and insightful suggestions. SN acknowledges support by the Science $\&$ Engineering Research Board, a statutory body of Department of Science $\&$  Technology (DST), Government of India (FILE NO. PDF/2018/002833). AC thanks FAPESP grants 2017/25651-5 and 2014/11156-4. We acknowledge the support of the Department of Atomic Energy, Government of India, under the project 12-R\&D-TFR-5.02-0700. The National Radio Astronomy Observatory is a facility of the National Science Foundation operated under cooperative agreement by Associated Universities, Inc. We thank the staff of the GMRT that made these observations possible. GMRT is run by the National Centre for Radio Astrophysics of the Tata Institute of Fundamental Research.

\begin{figure}
\centering{
\includegraphics[width=8cm]{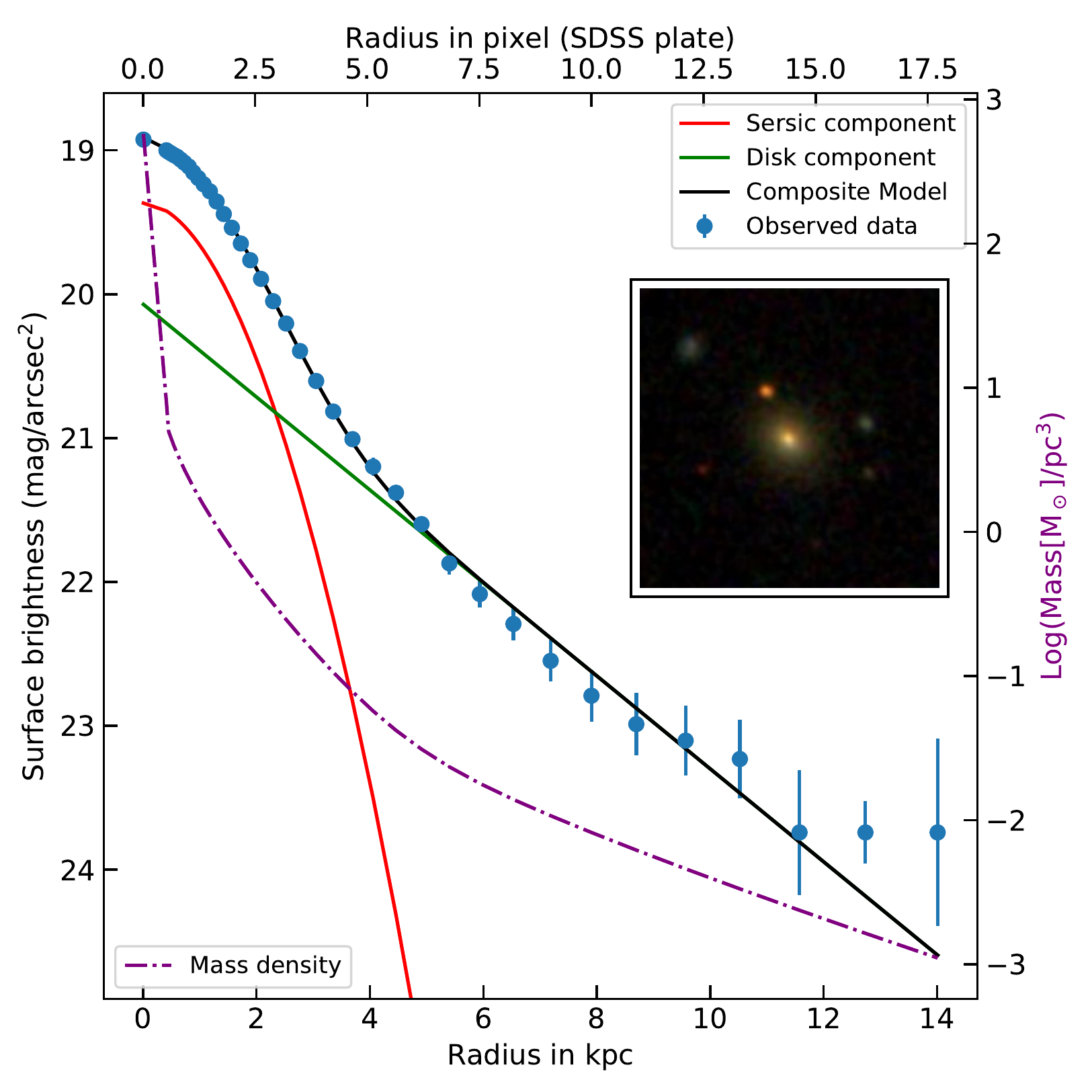}}
\caption{Surface brightness of J1328+2752 as a function of radius (blue points). Black curve shows the resulting model fit for SDSS r-band data from the combination of a Sersic and an exponential-disk profiles (red and green lines, respectively). Dotted-dashed purple line represents the derived stellar-density profile of J1328+2752, while the inset image shows the host galaxy in the r band.}
\label{Optical_fit}
\end{figure}

\appendix
{\section{Optical data for the host galaxy}\label{sec:SDSS}
The properties of the host galaxy J1328+2751 have been studied from Sloan Digital Sky Survey (SDSS) i and r band data. The concentric isophotes starting from the centroid of the galaxy are fitted for both bands to estimate the surface intensity distribution of the host. The outer most isophote extends up to $\sim$13 kpc in the galaxy frame. Both Sersic and exponential profiles are required to fit the host properly \citep{2005PASA...22..118G}. The bulge and disk decomposition are shown in Figure \ref{Optical_fit}. The disk component dominates beyond $\sim$2.5 kpc, whereas the inner portion is bulge dominated. For each isophotes stellar mass-to-light (M/L) ratios were calculated using \citet{2003ApJS..149..289B}. Figure \ref{Optical_fit} also shows the stellar-density distribution within the galaxy. We have assumed that the intensity of the centroid corresponds the minimum flux density coming from the sphere of influence of radius 6.3 parsec, where the SMBH pair is residing. This corresponds to  minimum stellar-density at the vicinity of BBH $\sim$575~M$_\odot$ parsec$^{-3}$ for this sphere.
The variation in surface brightness profile derived from the SDSS image of the optical host shows the presence of an extended disk like substructure. This is also a potential indicator of a gas-rich, unequal-mass merger \citep{2003ApJ...597..893N}.}

\begin{figure*}
\centering{
\includegraphics[width=16cm]{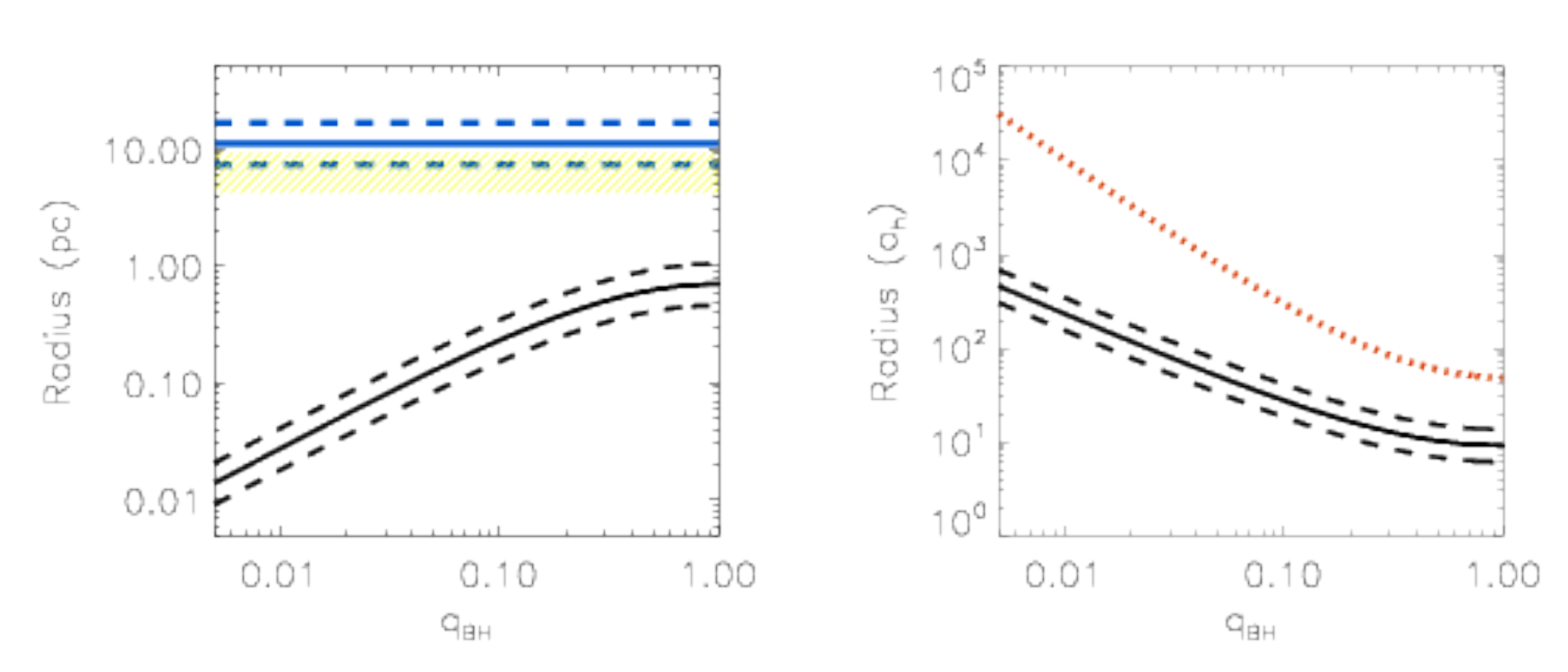}}
\caption{{\it Left panel:} Gravitational influence radius of the BBH (blue line) and the hardening radius (black line) in terms of $q_\mathrm{BH}$. Hatched yellow rectangle show the 1$\sigma$ limits for the separation of the two black holes derived from velocity offset of the double-peaked emission lines in the spectrum of J1328+2752. {\it Right panel:} Black hole separation divided by the hardening radius (black line) as a function of $q_\mathrm{BH}$. Red dotted line represents the condition $P_\mathrm{orb,obs}\sim t_\mathrm{h}$ considering $\rho_\ast = 575$ M$_\odot$ parsec$^{-3}$. Dashed lines refer to 1$\sigma$-level uncertainties.}
	\label{figureRb_Rh_vs_qBH_KeplerOrb}
\end{figure*}

\section{Is the putative BBH in J1328+2752 gravitationally bound?} \label{sec:SMBBSKepler}
	
	A BBH can be considered gravitationally bound if the separation between the black holes is smaller than the radius of a gravitational influence sphere of the BBH, $R_\mathrm{b}$, which is defined as \citep{2001ApJ...563...34M} 
	
	\begin{equation} \label{Rbound}
	R_\mathrm{b} = \frac{GM_\mathrm{tot}}{\sigma_\ast^2}.
	\end{equation}
	
	We show in the lower left panel of Figure \ref{figureRb_Rh_vs_qBH_KeplerOrb} the comparison of $R_\mathrm{b}$ and $d_\mathrm{BH}$, indicating that the two black holes are gravitationally bound for any value of $q_\mathrm{BH}$. In the same figure is also shown the behaviour of the hardness radius, $a_h$, in terms of of $q_\mathrm{BH}$, calculated from
	
	\begin{equation} \label{ahard}
	a_\mathrm{h} = \frac{GM_\mathrm{tot}}{4\sigma_\ast^2}\frac{q_\mathrm{BH}}{\left(1+q_\mathrm{BH}\right)^2}.
	\end{equation}
	
	Independent of the value of $q_\mathrm{BH}$, $d_\mathrm{BH}$ is always larger than $a_h$ at 1$\sigma$-level, implying that the BBH in J1328+2752 is not in the evolutionary hard phase, which it is thought to start when the binding energy per unit mass of the binary system exceeds the specific binding energy of surrounding stars \citep{1980Natur.287..307B}. Since the BBH in J1328+2752 is not in a hard stage, it is interesting to verify whether their orbits are Keplerian. Even though the real trajectories of the binary black holes during their evolution are complex, involving multiple gravitational interactions with stars, gas, and even additional black holes \citep{1980Natur.287..307B, 
2001ApJ...563...34M, 2003ApJ...596..860M, 
	2015ApJ...810...49V, 
	2018MNRAS.473.3410R, 
2020MNRAS.493.3676O}, 
	a comparison between the orbital period and the hardening timescale, $t_h$, can be used as an approximate probe for Keplerian condition \citep{2017ApJ...837..135R}. In this case, $P_\mathrm{orb,obs}\lesssim t_\mathrm{h}$ would imply a Keplerian orbit. Taking \citep{2017ApJ...837..135R} $t_\mathrm{h}=\sigma_\ast/4Ga\rho_\ast H$, where $a$ is the semi-major axis of the orbit, $\rho_\ast$ is the stellar mass density in the vicinity of the BBH, and $H\approx15\left[1+a/(3.5a_\mathrm{h})\right]^{-1}$, the dimensionless binary hardening
	rate \citep{2006ApJ...651..392S}, we can rewrite the Keplerian condition in terms of $a$ and $a_\mathrm{h}$ as
	
	\begin{equation} \label{KepOrbCond}
	\left(\frac{a}{a_\mathrm{h}}\right)^{5/2} - \left(\frac{32\sigma_\ast^6}{105\pi\rho_\ast G^3M_\mathrm{tot}^2}\right) \left(\frac{a}{a_\mathrm{h}}+\frac{7}{2}\right) \frac{\left(1+q_\mathrm{BH}\right)^5}{q_\mathrm{BH}^{5/2}}\lesssim 0.
	\end{equation}
	
	We show in the lower right Figure \ref{figureRb_Rh_vs_qBH_KeplerOrb} the numerical solution of equation (\ref{KepOrbCond}) for $\rho_\ast = 575$ M$_\odot$ parsec$^{-3}$, the minimum value for the stellar mass density close to the BBH derived from the optical data of the host galaxy of J1328+2752 (upper panel Figure \ref{figureRb_Rh_vs_qBH_KeplerOrb}). We also display in the same figure the behaviour of $d_\mathrm{BH}/a_\mathrm{h}$ in terms of $q_\mathrm{BH}$. We can note that the Keplerian condition is satisfied for any value of $q_\mathrm{BH}$, suggesting that orbital motions of the black holes in J1328+2752 are Keplerian.

\section{Observational constraints for the jet precession modelling} \label{sec:JetPrecJ1328}
\citet{Nandi17} found that the inner double structures (N and S components in Figure 1 in \citealt{Nandi17}) are misaligned by about $30^\circ$ from the axis of the outer diffuse structures (NW and SE components) in J1328+2752. Besides, the SE component has a structure similar to a lobe with an intense hot spot, indicating a possible interaction with the intergalactic medium at that location, which is not seen in the case of NW component. In the northern region, jet presents a pronounced bent that finishes close to the position angle of the (inner) N component. Similar behaviour is seen in the south part, but with an inverted bent in relation to the north. Those characteristics suggest that an interval corresponding to at least one precession period might be enough to generate the S-shaped morphology in J1328+2752. Therefore, the jet precession period must be shorter than about 52 Myr, the estimated upper limit for age of the radio structure observed in J1328+2752 \citep{2019MNRAS.486.5158N}. 
	
	On the other hand, a lower limit for the precession period can be determined assuming a non-accelerating motion for the jet. From the maximum apparent jet speed derived in the previous section and the projected physical length of the jet, we found that $P_\mathrm{prec,obs}\gtrsim 0.5$ Myr if the jet in J1328+2752 has precessed by approximately one precession period.
	
	The brightness distribution in our kiloparsec-scale radio maps shows jet features at sky position angles roughly between -50$^\circ$ and 20$^\circ$, which implies that $\eta_0$ must respect such range for a suitable jet precession model. Besides, we can use those sky position angles together with the instantaneous jet position angle predicted by our precession model \citet{2017ApJ...851L..39C} to estimate a maximum value for $\varphi_0$. This is obtained imposing $\phi_0\sim 86^\circ$ and varying $\varphi_0$ until the range of instantaneous jet position angle from the precession model begins to fall outside of the observed range for the jet position angles, which happens when $\varphi_0>35^\circ$ for J1328+2752.

\section{Implications for DDRG formation scenarios} \label{sec:DDRG}
Previous studies also have argued that DDRGs are the remnants of merged BBHs \citep{2003MNRAS.340..411L, 2004MNRAS.347.1357L, Komossa06}. The secondary black hole temporarily removes the inner parts of the accretion disk of the primary central black hole and the jet stops. Jet activity again starts after the outer parts of the accretion disk refill the inner parts of the disk of the primary black hole \citep{2003MNRAS.340..411L}. In the case of J1328$+$2752, the jet emission is misaligned for the two epochs. A nearby galaxy or a binary black hole merger may change the direction ejection axis. Therefore, these kinds of DDRGs are not remnants of merged BBHs, but rather are interacting systems.

\section{Gravitational Radiation}\label{sec:GRavRad}
Coalescence of binary SMBHs are a potential source of the intense gravitational wave radiation making these systems extremely useful for future gravitational wave experiments  \citep{2016PhRvL.116x1103A, 2017MNRAS.464.2301G}.
The natural gravitational wave frequency emitted by a BBH is

\begin{equation} 
 f_\mathrm{GW}^\mathrm{BBH}\approx \left(2\pi\right)^{-1}\sqrt{\frac{GM_\mathrm{tot}}{{d_\mathrm{BH}^{3}}}}
\end{equation}
For J1328$+$2752 the estimated $f_\mathrm{GW}^\mathrm{BBH}\sim2\times 10^{-13}$~Hz. This gravitational wave signal falls in the ultra low
frequency band, below the Laser Interferometer Space Antenna (LISA) detection limit. However, such signals are likely to interfere with the signals generated during the time of inflation. Sources like J1328+2752 and the radio galaxy 0402+379 are therefore direct probes of inflationary physics. 


\begin{thebibliography}{}
\expandafter\ifx\csname natexlab\endcsname\relax\def\natexlab#1{#1}\fi
\providecommand{\url}[1]{\href{#1}{#1}}
\providecommand{\dodoi}[1]{doi:~\href{http://doi.org/#1}{\nolinkurl{#1}}}
\providecommand{\doeprint}[1]{\href{http://ascl.net/#1}{\nolinkurl{http://ascl%
.net/#1}}}
\providecommand{\doarXiv}[1]{\href{https://arxiv.org/abs/#1}{\nolinkurl{https:%
//arxiv.org/abs/#1}}}

\bibitem[{{Abbott} {et~al.}(2016){Abbott}, {Abbott}, {Abbott}, {Abernathy},
  {Acernese}, {Ackley}, {Adams}, {Adams}, {Addesso}, {Adhikari}, \&
  et~al.}]{2016PhRvL.116x1103A}
{Abbott}, B.~P., {Abbott}, R., {Abbott}, T.~D., {et~al.} 2016, Physical Review
  Letters, 116, 241103, \dodoi{10.1103/PhysRevLett.116.241103}

\bibitem[{{Abraham}(2000)}]{2000A&A...355..915A}
{Abraham}, Z. 2000, \aap, 355, 915

\bibitem[{{Bardeen} \& {Petterson}(1975)}]{1975ApJ...195L..65B}
{Bardeen}, J.~M., \& {Petterson}, J.~A. 1975, \apjl, 195, L65,
  \dodoi{10.1086/181711}

\bibitem[{{Barker} \& {O'Connell}(1975)}]{1975ApJ...199L..25B}
{Barker}, B.~M., \& {O'Connell}, R.~F. 1975, \apjl, 199, L25,
  \dodoi{10.1086/181840}

\bibitem[{{Bate} {et~al.}(2000){Bate}, {Bonnell}, {Clarke}, {Lubow}, {Ogilvie},
  {Pringle}, \& {Tout}}]{2000MNRAS.317..773B}
{Bate}, M.~R., {Bonnell}, I.~A., {Clarke}, C.~J., {et~al.} 2000, \mnras, 317,
  773, \dodoi{10.1046/j.1365-8711.2000.03648.x}

\bibitem[{{Begelman} {et~al.}(1980){Begelman}, {Blandford}, \&
  {Rees}}]{1980Natur.287..307B}
{Begelman}, M.~C., {Blandford}, R.~D., \& {Rees}, M.~J. 1980, \nat, 287, 307,
  \dodoi{10.1038/287307a0}

\bibitem[{{Bell} {et~al.}(2003){Bell}, {McIntosh}, {Katz}, \&
  {Weinberg}}]{2003ApJS..149..289B}
{Bell}, E.~F., {McIntosh}, D.~H., {Katz}, N., \& {Weinberg}, M.~D. 2003, \apjs,
  149, 289, \dodoi{10.1086/378847}

\bibitem[{{Blanton} {et~al.}(2001){Blanton}, {Gregg}, {Helfand}, {Becker}, \&
  {Leighly}}]{Blanton01}
{Blanton}, E.~L., {Gregg}, M.~D., {Helfand}, D.~J., {Becker}, R.~H., \&
  {Leighly}, K.~M. 2001, \aj, 121, 2915, \dodoi{10.1086/321074}

\bibitem[{{Britzen} {et~al.}(2018){Britzen}, {Fendt}, {Witzel}, {Qian},
  {Pashchenko}, {Kurtanidze}, {Zajacek}, {Martinez}, {Karas}, {Aller}, {Aller},
  {Eckart}, {Nilsson}, {Ar{\'e}valo}, {Cuadra}, {Subroweit}, \&
  {Witzel}}]{2018MNRAS.478.3199B}
{Britzen}, S., {Fendt}, C., {Witzel}, G., {et~al.} 2018, \mnras, 478, 3199,
  \dodoi{10.1093/mnras/sty1026}

\bibitem[{{Caproni} \& {Abraham}(2004)}]{2004ApJ...602..625C}
{Caproni}, A., \& {Abraham}, Z. 2004, \apj, 602, 625, \dodoi{10.1086/381195}

\bibitem[{{Caproni} {et~al.}(2013){Caproni}, {Abraham}, \&
  {Monteiro}}]{2013MNRAS.428..280C}
{Caproni}, A., {Abraham}, Z., \& {Monteiro}, H. 2013, \mnras, 428, 280,
  \dodoi{10.1093/mnras/sts014}

\bibitem[{{Caproni} {et~al.}(2006{\natexlab{a}}){Caproni}, {Abraham}, \&
  {Mosquera Cuesta}}]{2006ApJ...638..120C}
{Caproni}, A., {Abraham}, Z., \& {Mosquera Cuesta}, H.~J. 2006{\natexlab{a}},
  \apj, 638, 120, \dodoi{10.1086/498684}

\bibitem[{{Caproni} {et~al.}(2017){Caproni}, {Abraham}, {Motter}, \&
  {Monteiro}}]{2017ApJ...851L..39C}
{Caproni}, A., {Abraham}, Z., {Motter}, J.~C., \& {Monteiro}, H. 2017, \apjl,
  851, L39, \dodoi{10.3847/2041-8213/aa9fea}

\bibitem[{{Caproni} {et~al.}(2006{\natexlab{b}}){Caproni}, {Livio}, {Abraham},
  \& {Mosquera Cuesta}}]{2006ApJ...653..112C}
{Caproni}, A., {Livio}, M., {Abraham}, Z., \& {Mosquera Cuesta}, H.~J.
  2006{\natexlab{b}}, \apj, 653, 112, \dodoi{10.1086/508508}

\bibitem[{{Caproni} {et~al.}(2009){Caproni}, {Monteiro}, \&
  {Abraham}}]{2009MNRAS.399.1415C}
{Caproni}, A., {Monteiro}, H., \& {Abraham}, Z. 2009, \mnras, 399, 1415,
  \dodoi{10.1111/j.1365-2966.2009.15361.x}

\bibitem[{{Comerford} {et~al.}(2012){Comerford}, {Gerke}, {Stern}, {Cooper},
  {Weiner}, {Newman}, {Madsen}, \& {Barrows}}]{2012ApJ...753...42C}
{Comerford}, J.~M., {Gerke}, B.~F., {Stern}, D., {et~al.} 2012, \apj, 753, 42,
  \dodoi{10.1088/0004-637X/753/1/42}

\bibitem[{{Comerford} {et~al.}(2009){Comerford}, {Gerke}, {Newman}, {Davis},
  {Yan}, {Cooper}, {Faber}, {Koo}, {Coil}, {Rosario}, \&
  {Dutton}}]{2009ApJ...698..956C}
{Comerford}, J.~M., {Gerke}, B.~F., {Newman}, J.~A., {et~al.} 2009, \apj, 698,
  956, \dodoi{10.1088/0004-637X/698/1/956}

\bibitem[{{Crenshaw} {et~al.}(2010){Crenshaw}, {Schmitt}, {Kraemer},
  {Mushotzky}, \& {Dunn}}]{2010ApJ...708..419C}
{Crenshaw}, D.~M., {Schmitt}, H.~R., {Kraemer}, S.~B., {Mushotzky}, R.~F., \&
  {Dunn}, J.~P. 2010, \apj, 708, 419, \dodoi{10.1088/0004-637X/708/1/419}

\bibitem[{{Deane} {et~al.}(2014){Deane}, {Paragi}, {Jarvis}, {Coriat},
  {Bernardi}, {Fender}, {Frey}, {Heywood}, {Kl{\"o}ckner}, {Grainge}, \&
  {Rumsey}}]{Deane14}
{Deane}, R.~P., {Paragi}, Z., {Jarvis}, M.~J., {et~al.} 2014, \nat, 511, 57,
  \dodoi{10.1038/nature13454}

\bibitem[{{Eggleton}(1983)}]{1983ApJ...268..368E}
{Eggleton}, P.~P. 1983, \apj, 268, 368, \dodoi{10.1086/160960}

\bibitem[{{Ekers} {et~al.}(1978){Ekers}, {Fanti}, {Lari}, \&
  {Parma}}]{1978Natur.276..588E}
{Ekers}, R.~D., {Fanti}, R., {Lari}, C., \& {Parma}, P. 1978, \nat, 276, 588,
  \dodoi{10.1038/276588a0}

\bibitem[{{Falceta-Gon{\c{c}}alves} {et~al.}(2010){Falceta-Gon{\c{c}}alves},
  {Caproni}, {Abraham}, {Teixeira}, \& {de Gouveia Dal
  Pino}}]{2010ApJ...713L..74F}
{Falceta-Gon{\c{c}}alves}, D., {Caproni}, A., {Abraham}, Z., {Teixeira}, D.~M.,
  \& {de Gouveia Dal Pino}, E.~M. 2010, \apjl, 713, L74,
  \dodoi{10.1088/2041-8205/713/1/L74}

\bibitem[{{Ferrarese} \& {Merritt}(2000)}]{Ferrarese00}
{Ferrarese}, L., \& {Merritt}, D. 2000, \apjl, 539, L9, \dodoi{10.1086/312838}

\bibitem[{{Fragile} {et~al.}(2007){Fragile}, {Blaes}, {Anninos}, \&
  {Salmonson}}]{2007ApJ...668..417F}
{Fragile}, P.~C., {Blaes}, O.~M., {Anninos}, P., \& {Salmonson}, J.~D. 2007,
  \apj, 668, 417, \dodoi{10.1086/521092}

\bibitem[{{Garon} {et~al.}(2019){Garon}, {Rudnick}, {Wong}, {Jones}, {Kim},
  {Andernach}, {Shabala}, {Kapi{\'n}ska}, {Norris}, {de Gasperin}, {Tate}, \&
  {Tang}}]{2019AJ....157..126G}
{Garon}, A.~F., {Rudnick}, L., {Wong}, O.~I., {et~al.} 2019, \aj, 157, 126,
  \dodoi{10.3847/1538-3881/aaff62}

\bibitem[{{Genzel} {et~al.}(1997){Genzel}, {Eckart}, {Ott}, \&
  {Eisenhauer}}]{Genzel97}
{Genzel}, R., {Eckart}, A., {Ott}, T., \& {Eisenhauer}, F. 1997, \mnras, 291,
  219, \dodoi{10.1093/mnras/291.1.219}

\bibitem[{{Gerosa} {et~al.}(2020){Gerosa}, {Rosotti}, \&
  {Barbieri}}]{2020MNRAS.496.3060G}
{Gerosa}, D., {Rosotti}, G., \& {Barbieri}, R. 2020, \mnras, 496, 3060,
  \dodoi{10.1093/mnras/staa1693}

\bibitem[{{Goodman}(2003)}]{2003MNRAS.339..937G}
{Goodman}, J. 2003, \mnras, 339, 937, \dodoi{10.1046/j.1365-8711.2003.06241.x}

\bibitem[{{Gower} {et~al.}(1982{\natexlab{a}}){Gower}, {Gregory}, {Unruh}, \&
  {Hutchings}}]{1982ApJ...262..478G}
{Gower}, A.~C., {Gregory}, P.~C., {Unruh}, W.~G., \& {Hutchings}, J.~B.
  1982{\natexlab{a}}, \apj, 262, 478, \dodoi{10.1086/160442}

\bibitem[{{Gower} {et~al.}(1982{\natexlab{b}}){Gower}, {Gregory}, {Unruh}, \&
  {Hutchings}}]{Gower82}
---. 1982{\natexlab{b}}, \apj, 262, 478, \dodoi{10.1086/160442}

\bibitem[{{Graham} \& {Driver}(2005)}]{2005PASA...22..118G}
{Graham}, A.~W., \& {Driver}, S.~P. 2005, \pasa, 22, 118,
  \dodoi{10.1071/AS05001}

\bibitem[{{Greenhill} {et~al.}(1996){Greenhill}, {Gwinn}, {Antonucci}, \&
  {Barvainis}}]{1996ApJ...472L..21G}
{Greenhill}, L.~J., {Gwinn}, C.~R., {Antonucci}, R., \& {Barvainis}, R. 1996,
  \apjl, 472, L21, \dodoi{10.1086/310346}

\bibitem[{{Gualandris} {et~al.}(2017){Gualandris}, {Read}, {Dehnen}, \&
  {Bortolas}}]{2017MNRAS.464.2301G}
{Gualandris}, A., {Read}, J.~I., {Dehnen}, W., \& {Bortolas}, E. 2017, \mnras,
  464, 2301, \dodoi{10.1093/mnras/stw2528}

\bibitem[{{Hardee}(1987)}]{Hardee87}
{Hardee}, P.~E. 1987, \apj, 318, 78, \dodoi{10.1086/165352}

\bibitem[{{Harms} {et~al.}(1994){Harms}, {Ford}, {Tsvetanov}, {Hartig},
  {Dressel}, {Kriss}, {Bohlin}, {Davidsen}, {Margon}, \& {Kochhar}}]{Harms94}
{Harms}, R.~J., {Ford}, H.~C., {Tsvetanov}, Z.~I., {et~al.} 1994, \apjl, 435,
  L35, \dodoi{10.1086/187588}

\bibitem[{{Herrnstein} {et~al.}(2005){Herrnstein}, {Moran}, {Greenhill}, \&
  {Trotter}}]{2005ApJ...629..719H}
{Herrnstein}, J.~R., {Moran}, J.~M., {Greenhill}, L.~J., \& {Trotter}, A.~S.
  2005, \apj, 629, 719, \dodoi{10.1086/431421}

\bibitem[{{Hutchings} {et~al.}(1988){Hutchings}, {Price}, \&
  {Gower}}]{1988ApJ...329..122H}
{Hutchings}, J.~B., {Price}, R., \& {Gower}, A.~C. 1988, \apj, 329, 122,
  \dodoi{10.1086/166363}

\bibitem[{{Kaspi} {et~al.}(2000){Kaspi}, {Smith}, {Netzer}, {Maoz}, {Jannuzi},
  \& {Giveon}}]{2000ApJ...533..631K}
{Kaspi}, S., {Smith}, P.~S., {Netzer}, H., {et~al.} 2000, \apj, 533, 631,
  \dodoi{10.1086/308704}

\bibitem[{{Katz}(1997)}]{1997ApJ...478..527K}
{Katz}, J.~I. 1997, \apj, 478, 527, \dodoi{10.1086/303811}

\bibitem[{{Kharb} {et~al.}(2015){Kharb}, {Das}, {Paragi}, {Subramanian}, \&
  {Chitta}}]{Kharb15}
{Kharb}, P., {Das}, M., {Paragi}, Z., {Subramanian}, S., \& {Chitta}, L.~P.
  2015, \apj, 799, 161, \dodoi{10.1088/0004-637X/799/2/161}

\bibitem[{{Kharb} {et~al.}(2017){Kharb}, {Lal}, \& {Merritt}}]{Kharb17}
{Kharb}, P., {Lal}, D.~V., \& {Merritt}, D. 2017, Nature Astronomy, 1, 727,
  \dodoi{10.1038/s41550-017-0256-4}

\bibitem[{{Kharb} {et~al.}(2020){Kharb}, {Lena}, {Paragi}, {Subramanian},
  {Vaddi}, {Das}, \& {Khatun}}]{Kharb2020}
{Kharb}, P., {Lena}, D., {Paragi}, Z., {et~al.} 2020, \apj, 890, 40,
  \dodoi{10.3847/1538-4357/ab6a09}

\bibitem[{{Kharb} {et~al.}(2006){Kharb}, {O'Dea}, {Baum}, {Colbert}, \&
  {Xu}}]{Kharb2006}
{Kharb}, P., {O'Dea}, C.~P., {Baum}, S.~A., {Colbert}, E.~J.~M., \& {Xu}, C.
  2006, \apj, 652, 177, \dodoi{10.1086/507945}

\bibitem[{{Kharb} {et~al.}(2014){Kharb}, {O'Dea}, {Baum}, {Hardcastle},
  {Dicken}, {Croston}, {Mingo}, \& {Noel-Storr}}]{Kharb14}
{Kharb}, P., {O'Dea}, C.~P., {Baum}, S.~A., {et~al.} 2014, \mnras, 440, 2976,
  \dodoi{10.1093/mnras/stu421}

\bibitem[{{Kharb} {et~al.}(2019){Kharb}, {Vaddi}, {Sebastian}, {Subramanian},
  {Das}, \& {Paragi}}]{Kharb2019}
{Kharb}, P., {Vaddi}, S., {Sebastian}, B., {et~al.} 2019, \apj, 871, 249,
  \dodoi{10.3847/1538-4357/aafad7}

\bibitem[{{King} \& {Pringle}(2007)}]{2007MNRAS.377L..25K}
{King}, A.~R., \& {Pringle}, J.~E. 2007, \mnras, 377, L25,
  \dodoi{10.1111/j.1745-3933.2007.00296.x}

\bibitem[{{Komossa}(2006)}]{Komossa06}
{Komossa}, S. 2006, \memsai, 77, 733

\bibitem[{{Krause} {et~al.}(2019){Krause}, {Shabala}, {Hardcastle}, {Bicknell},
  {B{\"o}hringer}, {Chon}, {Nawaz}, {Sarzi}, \& {Wagner}}]{2019MNRAS.482..240K}
{Krause}, M. G.~H., {Shabala}, S.~S., {Hardcastle}, M.~J., {et~al.} 2019,
  \mnras, 482, 240, \dodoi{10.1093/mnras/sty2558}

\bibitem[{{Kuo} {et~al.}(2011){Kuo}, {Braatz}, {Condon}, {Impellizzeri}, {Lo},
  {Zaw}, {Schenker}, {Henkel}, {Reid}, \& {Greene}}]{2011ApJ...727...20K}
{Kuo}, C.~Y., {Braatz}, J.~A., {Condon}, J.~J., {et~al.} 2011, \apj, 727, 20,
  \dodoi{10.1088/0004-637X/727/1/20}

\bibitem[{{Ku{\'z}micz} \& {Jamrozy}(2012)}]{2012MNRAS.426..851K}
{Ku{\'z}micz}, A., \& {Jamrozy}, M. 2012, \mnras, 426, 851,
  \dodoi{10.1111/j.1365-2966.2012.21576.x}

\bibitem[{{Lal} \& {Rao}(2007)}]{lal2007}
{Lal}, D.~V., \& {Rao}, A.~P. 2007, \mnras, 374, 1085,
  \dodoi{10.1111/j.1365-2966.2006.11225.x}

\bibitem[{{Larwood} {et~al.}(1996){Larwood}, {Nelson}, {Papaloizou}, \&
  {Terquem}}]{1996MNRAS.282..597L}
{Larwood}, J.~D., {Nelson}, R.~P., {Papaloizou}, J.~C.~B., \& {Terquem}, C.
  1996, \mnras, 282, 597, \dodoi{10.1093/mnras/282.2.597}

\bibitem[{{Lehnert} {et~al.}(1999){Lehnert}, {van Breugel}, {Heckman}, \&
  {Miley}}]{Lehnert99}
{Lehnert}, M.~D., {van Breugel}, W. J.~M., {Heckman}, T.~M., \& {Miley}, G.~K.
  1999, \apjs, 124, 11, \dodoi{10.1086/313252}

\bibitem[{{Lense} \& {Thirring}(1918)}]{1918PhyZ...19..156L}
{Lense}, J., \& {Thirring}, H. 1918, Physikalische Zeitschrift, 19, 156

\bibitem[{{Lind} \& {Blandford}(1985)}]{1985ApJ...295..358L}
{Lind}, K.~R., \& {Blandford}, R.~D. 1985, \apj, 295, 358,
  \dodoi{10.1086/163380}

\bibitem[{{Liska} {et~al.}(2018){Liska}, {Hesp}, {Tchekhovskoy}, {Ingram}, {van
  der Klis}, \& {Markoff}}]{2018MNRAS.474L..81L}
{Liska}, M., {Hesp}, C., {Tchekhovskoy}, A., {et~al.} 2018, \mnras, 474, L81,
  \dodoi{10.1093/mnrasl/slx174}

\bibitem[{{Liu}(2004)}]{2004MNRAS.347.1357L}
{Liu}, F.~K. 2004, \mnras, 347, 1357, \dodoi{10.1111/j.1365-2966.2004.07325.x}

\bibitem[{{Liu} {et~al.}(2003){Liu}, {Wu}, \& {Cao}}]{2003MNRAS.340..411L}
{Liu}, F.~K., {Wu}, X.-B., \& {Cao}, S.~L. 2003, \mnras, 340, 411,
  \dodoi{10.1046/j.1365-8711.2003.06235.x}

\bibitem[{{Lodato} \& {Pringle}(2006)}]{2006MNRAS.368.1196L}
{Lodato}, G., \& {Pringle}, J.~E. 2006, \mnras, 368, 1196,
  \dodoi{10.1111/j.1365-2966.2006.10194.x}

\bibitem[{{Lodato} \& {Pringle}(2007)}]{2007MNRAS.381.1287L}
---. 2007, \mnras, 381, 1287, \dodoi{10.1111/j.1365-2966.2007.12332.x}

\bibitem[{{Martin} {et~al.}(2007){Martin}, {Pringle}, \&
  {Tout}}]{2007MNRAS.381.1617M}
{Martin}, R.~G., {Pringle}, J.~E., \& {Tout}, C.~A. 2007, \mnras, 381, 1617,
  \dodoi{10.1111/j.1365-2966.2007.12349.x}

\bibitem[{{Milosavljevi{\'c}} \& {Merritt}(2001)}]{2001ApJ...563...34M}
{Milosavljevi{\'c}}, M., \& {Merritt}, D. 2001, \apj, 563, 34,
  \dodoi{10.1086/323830}

\bibitem[{{Milosavljevi{\'c}} \& {Merritt}(2003)}]{2003ApJ...596..860M}
---. 2003, \apj, 596, 860, \dodoi{10.1086/378086}

\bibitem[{{Miyoshi} {et~al.}(1995){Miyoshi}, {Moran}, {Herrnstein},
  {Greenhill}, {Nakai}, {Diamond}, \& {Inoue}}]{1995Natur.373..127M}
{Miyoshi}, M., {Moran}, J., {Herrnstein}, J., {et~al.} 1995, \nat, 373, 127,
  \dodoi{10.1038/373127a0}

\bibitem[{{Mukherjee} {et~al.}(2020){Mukherjee}, {Bodo}, {Mignone}, {Rossi}, \&
  {Vaidya}}]{2020MNRAS.499..681M}
{Mukherjee}, D., {Bodo}, G., {Mignone}, A., {Rossi}, P., \& {Vaidya}, B. 2020,
  \mnras, 499, 681, \dodoi{10.1093/mnras/staa2934}

\bibitem[{{Naab} \& {Burkert}(2003)}]{2003ApJ...597..893N}
{Naab}, T., \& {Burkert}, A. 2003, \apj, 597, 893, \dodoi{10.1086/378581}

\bibitem[{{Naab} \& {Ostriker}(2017)}]{2017ARA&A..55...59N}
{Naab}, T., \& {Ostriker}, J.~P. 2017, \araa, 55, 59,
  \dodoi{10.1146/annurev-astro-081913-040019}

\bibitem[{{Nandi} {et~al.}(2017){Nandi}, {Jamrozy}, {Roy}, {Larsson}, {Saikia},
  {Baes}, \& {Singh}}]{Nandi17}
{Nandi}, S., {Jamrozy}, M., {Roy}, R., {et~al.} 2017, \mnras, 467, L56,
  \dodoi{10.1093/mnrasl/slw256}

\bibitem[{{Nandi} {et~al.}(2019){Nandi}, {Saikia}, {Roy}, {Dabhade},
  {Wadadekar}, {Larsson}, {Baes}, {Chandola}, \& {Singh}}]{2019MNRAS.486.5158N}
{Nandi}, S., {Saikia}, D.~J., {Roy}, R., {et~al.} 2019, \mnras, 486, 5158,
  \dodoi{10.1093/mnras/stz1184}

\bibitem[{{Natarajan} \& {Pringle}(1998)}]{1998ApJ...506L..97N}
{Natarajan}, P., \& {Pringle}, J.~E. 1998, \apjl, 506, L97,
  \dodoi{10.1086/311658}

\bibitem[{{Nawaz} {et~al.}(2016){Nawaz}, {Bicknell}, {Wagner}, {Sutherland },
  \& {McNamara}}]{2016MNRAS.458..802N}
{Nawaz}, M.~A., {Bicknell}, G.~V., {Wagner}, A.~Y., {Sutherland }, R.~S., \&
  {McNamara}, B.~R. 2016, \mnras, 458, 802, \dodoi{10.1093/mnras/stw330}

\bibitem[{{Nealon} {et~al.}(2015){Nealon}, {Price}, \&
  {Nixon}}]{2015MNRAS.448.1526N}
{Nealon}, R., {Price}, D.~J., \& {Nixon}, C.~J. 2015, \mnras, 448, 1526,
  \dodoi{10.1093/mnras/stv014}

\bibitem[{{Nelson} \& {Papaloizou}(2000)}]{2000MNRAS.315..570N}
{Nelson}, R.~P., \& {Papaloizou}, J. C.~B. 2000, \mnras, 315, 570,
  \dodoi{10.1046/j.1365-8711.2000.03478.x}

\bibitem[{{Ogilvie}(1999)}]{1999MNRAS.304..557O}
{Ogilvie}, G.~I. 1999, \mnras, 304, 557,
  \dodoi{10.1046/j.1365-8711.1999.02340.x}

\bibitem[{{Ogiya} {et~al.}(2020){Ogiya}, {Hahn}, {Mingarelli}, \&
  {Volonteri}}]{2020MNRAS.493.3676O}
{Ogiya}, G., {Hahn}, O., {Mingarelli}, C. M.~F., \& {Volonteri}, M. 2020,
  \mnras, 493, 3676, \dodoi{10.1093/mnras/staa444}

\bibitem[{{Palenzuela} {et~al.}(2010){Palenzuela}, {Lehner}, \&
  {Liebling}}]{palenzula2010}
{Palenzuela}, C., {Lehner}, L., \& {Liebling}, S.~L. 2010, Science, 329, 927,
  \dodoi{10.1126/science.1191766}

\bibitem[{{Papaloizou} \& {Pringle}(1977)}]{1977MNRAS.181..441P}
{Papaloizou}, J., \& {Pringle}, J.~E. 1977, \mnras, 181, 441,
  \dodoi{10.1093/mnras/181.3.441}

\bibitem[{{Parma} {et~al.}(1985){Parma}, {Ekers}, \&
  {Fanti}}]{1985A&AS...59..511P}
{Parma}, P., {Ekers}, R.~D., \& {Fanti}, R. 1985, \aaps, 59, 511

\bibitem[{{Pringle}(1981)}]{1981ARA&A..19..137P}
{Pringle}, J.~E. 1981, \araa, 19, 137,
  \dodoi{10.1146/annurev.aa.19.090181.001033}

\bibitem[{{Rasskazov} \& {Merritt}(2017)}]{2017ApJ...837..135R}
{Rasskazov}, A., \& {Merritt}, D. 2017, \apj, 837, 135,
  \dodoi{10.3847/1538-4357/aa6188}

\bibitem[{{Rawes} {et~al.}(2018){Rawes}, {Birkinshaw}, \&
  {Worrall}}]{2018MNRAS.480.3644R}
{Rawes}, J., {Birkinshaw}, M., \& {Worrall}, D.~M. 2018, \mnras, 480, 3644,
  \dodoi{10.1093/mnras/sty2074}

\bibitem[{{Rodriguez} {et~al.}(2006){Rodriguez}, {Taylor}, {Zavala}, {Peck},
  {Pollack}, \& {Romani}}]{Rodriguez06}
{Rodriguez}, C., {Taylor}, G.~B., {Zavala}, R.~T., {et~al.} 2006, \apj, 646,
  49, \dodoi{10.1086/504825}

\bibitem[{{Romero} {et~al.}(2000){Romero}, {Chajet}, {Abraham}, \&
  {Fan}}]{2000A&A...360...57R}
{Romero}, G.~E., {Chajet}, L., {Abraham}, Z., \& {Fan}, J.~H. 2000, \aap, 360,
  57

\bibitem[{{Rosario} {et~al.}(2011){Rosario}, {McGurk}, {Max}, {Shields},
  {Smith}, \& {Ammons}}]{2011ApJ...739...44R}
{Rosario}, D.~J., {McGurk}, R.~C., {Max}, C.~E., {et~al.} 2011, \apj, 739, 44,
  \dodoi{10.1088/0004-637X/739/1/44}

\bibitem[{{Rubinur} {et~al.}(2018){Rubinur}, {Das}, \& {Kharb}}]{Rubinur2018}
{Rubinur}, K., {Das}, M., \& {Kharb}, P. 2018, Journal of Astrophysics and
  Astronomy, 39, 8, \dodoi{10.1007/s12036-018-9512-y}

\bibitem[{{Rubinur} {et~al.}(2019){Rubinur}, {Das}, \&
  {Kharb}}]{2019MNRAS.484.4933R}
---. 2019, \mnras, 484, 4933, \dodoi{10.1093/mnras/stz334}

\bibitem[{{Rubinur} {et~al.}(2017){Rubinur}, {Das}, {Kharb}, \&
  {Honey}}]{Rubinur2017}
{Rubinur}, K., {Das}, M., {Kharb}, P., \& {Honey}, M. 2017, \mnras, 465, 4772,
  \dodoi{10.1093/mnras/stw2981}

\bibitem[{{Ryu} {et~al.}(2018){Ryu}, {Perna}, {Haiman}, {Ostriker}, \&
  {Stone}}]{2018MNRAS.473.3410R}
{Ryu}, T., {Perna}, R., {Haiman}, Z., {Ostriker}, J.~P., \& {Stone}, N.~C.
  2018, \mnras, 473, 3410, \dodoi{10.1093/mnras/stx2524}

\bibitem[{{Saripalli} \& {Subrahmanyan}(2009)}]{saripalli2009}
{Saripalli}, L., \& {Subrahmanyan}, R. 2009, \apj, 695, 156,
  \dodoi{10.1088/0004-637X/695/1/156}

\bibitem[{{Scheuer} \& {Feiler}(1996)}]{1996MNRAS.282..291S}
{Scheuer}, P.~A.~G., \& {Feiler}, R. 1996, \mnras, 282, 291,
  \dodoi{10.1093/mnras/282.1.291}

\bibitem[{{Schoenmakers} {et~al.}(2000){Schoenmakers}, {de Bruyn},
  {R{\"o}ttgering}, {van der Laan}, \& {Kaiser}}]{2000MNRAS.315..371S}
{Schoenmakers}, A.~P., {de Bruyn}, A.~G., {R{\"o}ttgering}, H.~J.~A., {van der
  Laan}, H., \& {Kaiser}, C.~R. 2000, \mnras, 315, 371,
  \dodoi{10.1046/j.1365-8711.2000.03430.x}

\bibitem[{{Sebastian} {et~al.}(2019){Sebastian}, {Kharb}, {O'Dea}, {Gallimore},
  \& {Baum}}]{sebastian2019}
{Sebastian}, B., {Kharb}, P., {O'Dea}, C.~P., {Gallimore}, J.~F., \& {Baum},
  S.~A. 2019, \mnras, 490, L26, \dodoi{10.1093/mnrasl/slz136}

\bibitem[{{Sesana} {et~al.}(2006){Sesana}, {Haardt}, \&
  {Madau}}]{2006ApJ...651..392S}
{Sesana}, A., {Haardt}, F., \& {Madau}, P. 2006, \apj, 651, 392,
  \dodoi{10.1086/507596}

\bibitem[{{Shakura} \& {Sunyaev}(1973)}]{1973A&A....24..337S}
{Shakura}, N.~I., \& {Sunyaev}, R.~A. 1973, \aap, 500, 33

\bibitem[{{Sillanpaa} {et~al.}(1988){Sillanpaa}, {Haarala}, {Valtonen},
  {Sundelius}, \& {Byrd}}]{1988ApJ...325..628S}
{Sillanpaa}, A., {Haarala}, S., {Valtonen}, M.~J., {Sundelius}, B., \& {Byrd},
  G.~G. 1988, \apj, 325, 628, \dodoi{10.1086/166033}

\bibitem[{{Sirko} \& {Goodman}(2003)}]{2003MNRAS.341..501S}
{Sirko}, E., \& {Goodman}, J. 2003, \mnras, 341, 501,
  \dodoi{10.1046/j.1365-8711.2003.06431.x}

\bibitem[{{Smith} {et~al.}(2012){Smith}, {Shields}, {Salviander}, {Stevens}, \&
  {Rosario}}]{2012ApJ...752...63S}
{Smith}, K.~L., {Shields}, G.~A., {Salviander}, S., {Stevens}, A.~C., \&
  {Rosario}, D.~J. 2012, \apj, 752, 63, \dodoi{10.1088/0004-637X/752/1/63}

\bibitem[{{Taylor} \& {Vermeulen}(1997)}]{1997ApJ...485L...9T}
{Taylor}, G.~B., \& {Vermeulen}, R.~C. 1997, \apjl, 485, L9,
  \dodoi{10.1086/310800}

\bibitem[{{Toomre} \& {Toomre}(1972)}]{1972ApJ...178..623T}
{Toomre}, A., \& {Toomre}, J. 1972, \apj, 178, 623, \dodoi{10.1086/151823}

\bibitem[{Ulvestad {et~al.}(2005)Ulvestad, Antonucci, \&
  Barvainis}]{Ulvestad05}
Ulvestad, J.~S., Antonucci, R. R.~J., \& Barvainis, R. 2005, The Astrophysical
  Journal, 621, 123, \dodoi{10.1086/427426}

\bibitem[{{Vasiliev} {et~al.}(2015){Vasiliev}, {Antonini}, \&
  {Merritt}}]{2015ApJ...810...49V}
{Vasiliev}, E., {Antonini}, F., \& {Merritt}, D. 2015, \apj, 810, 49,
  \dodoi{10.1088/0004-637X/810/1/49}

\bibitem[{{Wang} {et~al.}(2020){Wang}, {Hammer}, {Rejkuba}, {Crnojevi{\'c}}, \&
  {Yang}}]{2020MNRAS.498.2766W}
{Wang}, J., {Hammer}, F., {Rejkuba}, M., {Crnojevi{\'c}}, D., \& {Yang}, Y.
  2020, \mnras, 498, 2766, \dodoi{10.1093/mnras/staa2508}

\bibitem[{{Wilkins}(1972)}]{1972PhRvD...5..814W}
{Wilkins}, D.~C. 1972, \prd, 5, 814, \dodoi{10.1103/PhysRevD.5.814}

\end{thebibliography}

\end{document}